\documentclass{emulateapj}
\usepackage{txfonts}
\usepackage{graphicx}
\usepackage{subfigure}
\usepackage{multirow}
\usepackage{url}

\shorttitle{PTF System and Performance}
\shortauthors{N.~Law et al.}

\begin{document}

\title{The Palomar Transient Factory: System Overview, Performance and First Results}

\author{
Nicholas~M. Law\altaffilmark{1}, 
Shrinivas~R. Kulkarni\altaffilmark{1}, 
Richard~G. Dekany\altaffilmark{1}, 
Eran~O. Ofek\altaffilmark{1}, 
Robert~M. Quimby\altaffilmark{1}, 
Peter~E. Nugent\altaffilmark{2}, 
Jason Surace\altaffilmark{3},  
Carl~C. Grillmair\altaffilmark{3},  
Joshua~S. Bloom\altaffilmark{4},
Mansi~M. Kasliwal\altaffilmark{1}, 
Lars Bildsten\altaffilmark{5},  
Tim Brown\altaffilmark{8},
S.~Bradley Cenko\altaffilmark{4}, 
David Ciardi\altaffilmark{3}, 
Ernest Croner\altaffilmark{1}, 
S.~George Djorgovski\altaffilmark{1}, 
Julian van Eyken\altaffilmark{3}, 
Alexei~V. Filippenko\altaffilmark{4},  
Derek~B. Fox\altaffilmark{6}, 
Avishay Gal-Yam\altaffilmark{7}, 
David Hale\altaffilmark{1}, 
Nouhad Hamam\altaffilmark{3},
George Helou\altaffilmark{3}, 
John Henning\altaffilmark{1}, 
D.~Andrew Howell\altaffilmark{8,12},  
Janet Jacobsen\altaffilmark{2},
Russ Laher\altaffilmark{3},  
Sean Mattingly\altaffilmark{3},  
Dan McKenna\altaffilmark{1}, 
Andrew Pickles\altaffilmark{8},
Dovi Poznanski\altaffilmark{2,4}
Gustavo Rahmer\altaffilmark{1}, 
Arne Rau\altaffilmark{1,9},
Wayne Rosing\altaffilmark{8},
Michael~Shara\altaffilmark{10},  
Roger Smith\altaffilmark{1}, 
Dan Starr\altaffilmark{4,8},
Mark Sullivan\altaffilmark{11},
Viswa Velur\altaffilmark{1}, 
Richard Walters\altaffilmark{1},
Jeff Zolkower\altaffilmark{1}
}

\affil{$^{1}$ Caltech Optical Observatories, MS 105-24, California Institute of Technology, Pasadena, CA 91125, USA}
\affil{$^{2}$ Lawrence Berkeley National Laboratory, Berkeley, CA 94720, USA}
\affil{$^{3}$ Infrared Processing and Analysis Center, California Institute of Technology, MS 100-22, Pasadena, CA 91125, USA}
\affil{$^{4}$ Department of Astronomy, University of California, Berkeley, CA 94720-3411, USA}
\affil{$^{5}$ Kavli Institute for Theoretical Physics and Department of Physics, University of California, Santa Barbara, CA 93106, USA}
\affil{$^{6}$ Department of Astronomy and Astrophysics, Pennsylvania State University, University Park, PA 16802, USA}
\affil{$^{7}$ Benoziyo Center for Astrophysics, Weizmann Institute of Science, 76100 Rehovot, Israel}
\affil{$^{8}$ Las Cumbres Observatory Global Telescope Network, 6740 Cortona Dr. Santa Barbara, CA 93117, USA}
\affil{$^{9}$ Max-Planck Institute for Extra-Terrestrial Physics, 85748 Garching, Germany}
\affil{$^{10}$ Department of Astrophysics, American Museum of Natural History, New York, NY 10024, USA}
\affil{$^{11}$ Oxford Astrophysics, Department of Physics, Denys Wilkinson Building, Keble Road, Oxford, OX1 3RH, UK}
\affil{$^{12}$ Department of Physics, University of California, Santa Barbara, CA 93106, USA}

\email{nlaw@astro.caltech.edu}

\begin{abstract}
The Palomar Transient Factory (PTF) is a fully-automated, wide-field survey aimed at a systematic exploration of the optical transient sky. The transient survey is performed using a new 8.1 square degree camera installed on the 48-inch Samuel Oschin telescope at Palomar Observatory; colors and light curves for detected transients are obtained with the automated Palomar 60-inch telescope. PTF uses eighty percent of the 1.2-m and fifty percent of the 1.5-m telescope time. With an exposure of 60-s the survey reaches a depth of $\rm{m_{g^\prime}}$$\approx$21.3 and $\rm{m_R}$$\approx$20.6 (5$\sigma$, median seeing). Four major experiments are planned for the five-year project: 1) a 5-day cadence supernova search; 2) a rapid transient search with cadences between 90 seconds and 1 day; 3) a search for eclipsing binaries and transiting planets in Orion; and 4) a 3$\pi$ sr deep H-alpha survey. PTF provides automatic, realtime transient classification and followup, as well as a database including every source detected in each frame. This paper summarizes the PTF project, including several months of on-sky performance tests of the new survey camera, the observing plans and the data reduction strategy. We conclude by detailing the first 51 PTF optical transient detections, found in commissioning data.
\end{abstract}

\keywords{Astronomical Instrumentation, Data Analysis and Techniques, Supernovae, Quasars and Active Galactic Nuclei, Stars}

\maketitle

\section{Introduction}

The Palomar Transient Factory (PTF) is a comprehensive transient detection system including a wide-field survey camera, an automated realtime data reduction pipeline, a dedicated photometric follow up telescope, and a full archive of all detected sources. The survey camera achieved first light on 13 Dec 2008; the project is planned to complete commissioning in June 2009.

This paper describes the technical aspects of the PTF project. An accompanying paper (Rau et al. 2009) describes the science planned for PTF in detail; subsequent papers will discuss the various PTF pipelines in more detail as well as the results from the PTF surveys.

The transient detection survey component of PTF is performed at the automated Palomar Samuel Oschin 48-inch telescope (P48); candidate transients are photometrically followed up at the automated Palomar 60-inch telescope (P60). This dual-telescope approach allows both high survey throughput and a very flexible followup program.

PTF's survey camera is based on the CFH12K mosaic camera formerly at the Canada-France-Hawaii Telescope, newly mounted on the P48 telescope at Palomar Observatory, California (and called "the PTF Survey Camera"). The camera has 101 megapixels, 1-arcsec sampling and a 8.1 square-degree field of view. Observations are mainly performed in one of two broad-band filters (Mould-R, SDSS-$\rm{g^\prime}$). Under median seeing conditions (1.1 arcsec at Palomar) the camera achieves 2.0 arcsec FWHM images, and reaches 5$\sigma$ limiting AB magnitudes of  $\rm{m_{g^{\prime}}}$$\approx$21.3 and $\rm{m_R}$$\approx$20.6 in 60-second exposures.

\begin{figure}
  \centering
  \resizebox{\columnwidth}{!}
   {
	\includegraphics[angle=0]{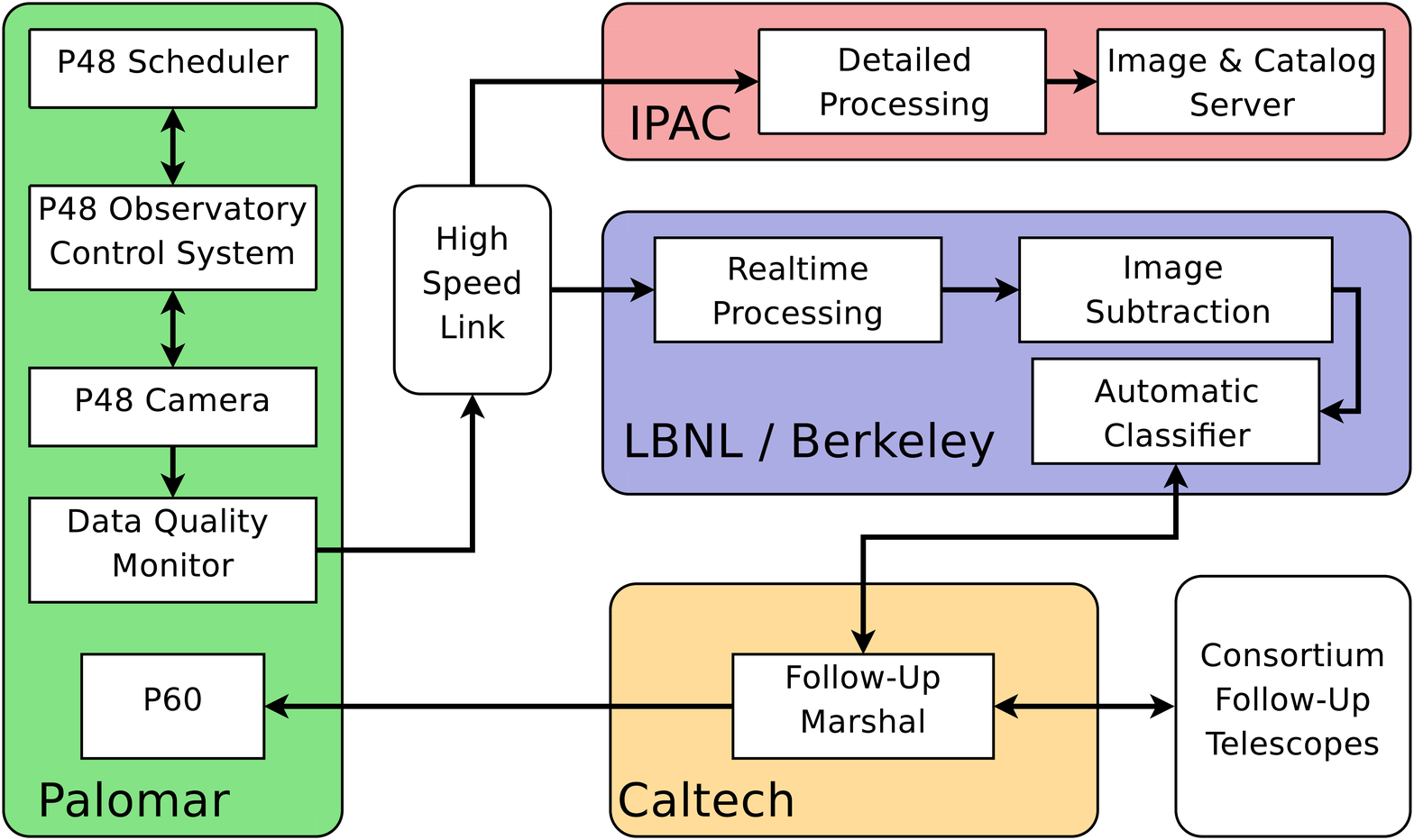}
   }
   \caption{An overview of the PTF project data flow.}
   \label{FIG:dataflow} 
\end{figure}

Data taken with the camera are transferred to two automated reduction pipelines (Figure \ref{FIG:dataflow}). A near-realtime image subtraction pipeline is run at Lawrence Berkeley National Laboratory (LBNL) and has the goal of identifying optical transients within minutes of images being taken. The output of this pipeline is sent to UC Berkeley where a source classifier determines a set of probabilistic statements about the scientific classification of the transients based on all available time-series and context data.

On few-day timescales the images are also ingested into a database at the Infrared Processing and Analysis Center (IPAC). Each incoming frame is calibrated and searched for objects, before the detections are merged into a database which is comprehensive, made public after an 18 month proprietary period and can be queried for all detections in an area of the sky.

Followup of detected transients is a vital component of successful transient surveys. The P60 photometric followup telescope automatically generates colors and light curves for interesting transients detected using P48. The PTF collaboration also leverages a further 15 telescopes for photometric and spectroscopic followup. An automated system will collate detections from the Berkeley classification engine, make them available to the various follow up facilities, coordinate the observations, and report on the results.

This paper is organized as follows: in Section \ref{sec:survey_system} we describe the new PTF survey camera and automated observing system, and detail the on-sky performance of the system. Section \ref{sec:strategy} discusses the initial PTF observing strategy. Section \ref{sec:pipelines} describes the automated data reduction pipeline and transient classification system, and Section \ref{sec:followup} details the PTF followup systems. In Section \ref{sec:sn} we conclude by describing PTF's first confirmed optical transient detections.

\section{The PTF Survey System: P48 + the PTF Survey Camera}
\label{sec:survey_system}

The P48 telescope is a wide-field Schmidt with a 48-inch aperture, a glass corrector plate and a 72-inch (f/2.5) mirror \citep{Harrington1952}. The telescope performed both Palomar Sky Surveys (POSS I and POSS II; \citet{Reid_1991}), and recently completed the Palomar-Quest digital synoptic sky survey \citep{Djorgovski_2008}, before the start of PTF modifications in autumn 2008.

\begin{figure*}
  \centering
  \resizebox{0.8\textwidth}{!}
   {
	\includegraphics[angle=0]{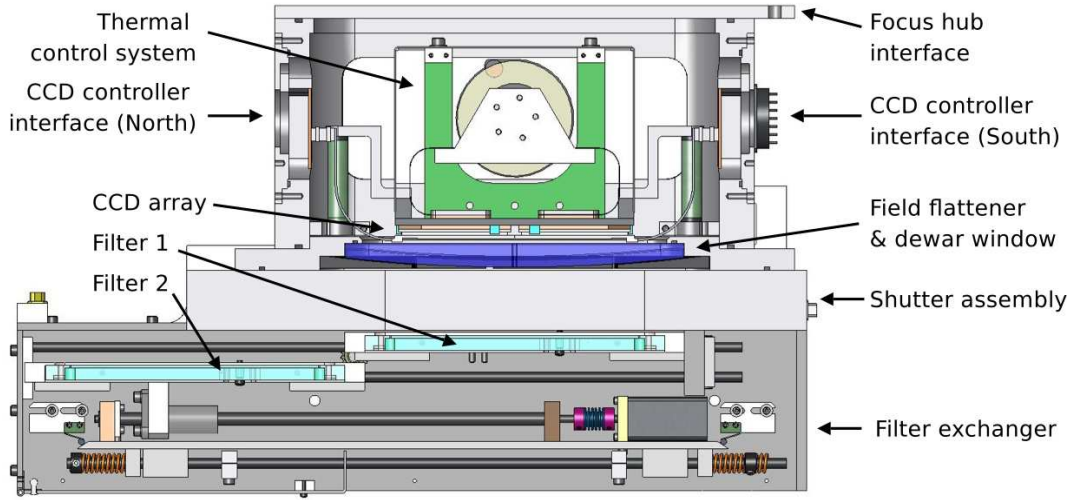}
   }
   \caption{Cut-away diagram of the PTF survey camera after modifications for use on the Palomar 48" telescope. The diagram is orientated so downwards is in the direction of the P48 primary mirror when mounted in the telescope. From bottom up, the components are: the filter exchanger, shutter, and the dewar assembly including the CCD window and CCD array. \vspace{0.3cm}}
   \label{fig:cutaway_camera}
\end{figure*}

The PTF survey camera (Figure \ref{fig:cutaway_camera}) is based on the CFH12K camera \citep{Cuillandre2000}. The camera was extensively re-engineered by Caltech Optical Observatories for faster readout, closed-cycle thermal control, and robust survey operation. The CCD focal plane was untouched but the rest of the camera was modified for reliable and low-cost operation on the P48. Primary requirements for the design, in addition to the mechanical and optical modifications required for operation on P48, were to reduce the operation cost of the camera, to minimize the beam obstruction, and to increase the readout speed for PTF operations.

The P48 telescope optics are significantly faster than the CFHT, leading to stringent requirements on the camera's CCD array flatness and optical quality. The low-operation-cost requirement led to swapping the camera cooling system from a $\rm{LN_2}$ system to a CryoTiger closed-cycle cooler. The camera was also upgraded with a new precision shutter and filter-changer assembly. The telescope was refurbished for operation with PTF and a new queue-scheduling automated observatory control system was implemented.

We summarize here the upgrades and provide on-sky performance test results for the PTF survey camera on P48; more detail on the camera engineering is provided in \citet{Rahmer2008}. Table \ref{tab:camera_specs} summarizes the survey and P60 followup system specifications.

\begin{table}
\caption{The specifications of the PTF survey system}
\label{tab:camera_specs}

\begin{small}
\begin{tabular}{ll}

\multicolumn{2}{l}{\bf P48 survey characteristics}\\
\hline
Telescope     & Palomar 48-inch (1.2m) Samuel Oschin \\
Camera field dimensions & 3.50 $\times$ 2.31 degrees\\
Camera field of view        & 8.07 square degrees \\
Light sensitive area        & 7.26 square degrees \\
Plate scale             &  1.01 arcsec / pixel\\
Efficiency           & 66\% open-shutter (slew during readout)\\
Sensitivity (median)          & $\rm{m_R}$$\approx$20.6 in 60 s, 5$\sigma$\\
                                              & $\rm{m_{g^\prime}}$$\approx$21.3 in 60 s, 5$\sigma$ \\
Image quality        & 2.0 arcsec FWHM in median seeing \\
Filters              & $\rm{g^\prime}$ \& Mould-R; other bands available\\
\hline \vspace{0.25cm} \\
\multicolumn{2}{l}{\bf P48 survey camera CCD array }\\
\hline
Component CCDs       & 12 CCDs; 1 non-functional\\
CCD specs            & 2K$\times$4K MIT/LL 3-edge butted CCDs\\
Array Leveling       & Flat to within 20 microns \\
Pixels                      & 15 microns/pixel; 100\% filling factor\\
Chip gaps              & Median 35 pixels (35 arcsec) \\
Readout noise        & $<$ 20 $\rm{e^-}$\\
Readout speed        & 30 seconds, entire 100 MPix array\\
Linearity              & better than 0.5\% up to 60K ADUs \\
Optical distortion           & maximum 7\arcsec at array corners \\
                             & compared to flat grid\\
\hline \vspace{0.25cm} \\
\multicolumn{2}{l}{\bf P60 followup camera specifications }\\
\hline
Telescope & Palomar Observatory 60-inch (1.5m) \\
CCD specs            & 2K$\times$2K CCD\\
Plate scale                     & 0.38 arcsec / pixel\\
Readout noise        &  5 $\rm{e^-}$ (amp 1); 8 $\rm{e^-}$ (amp 2) \\
Readout speed        & 25 seconds (full frame)\\
Linearity              & better than 1\% up to 20K ADUs \\
Sensitivity (median)          & $\rm{m_g^\prime}$$\approx$21.6 in 120 s, 5$\sigma$\\
                                              & $\rm{m_{r^\prime}}$$\approx$21.3 in 120 s, 5$\sigma$ \\
                                              & $\rm{m_{i^\prime}}$$\approx$21.1 in 120 s, 5$\sigma$ \\
                                              & $\rm{m_{z^\prime}}$$\approx$20.0 in 120 s, 5$\sigma$ \\
\hline \vspace{0.1cm} \\
\end{tabular}
\end{small}

\end{table}

\begin{figure}
  \centering
  \resizebox{\columnwidth}{!}
   {
	\includegraphics[angle=0]{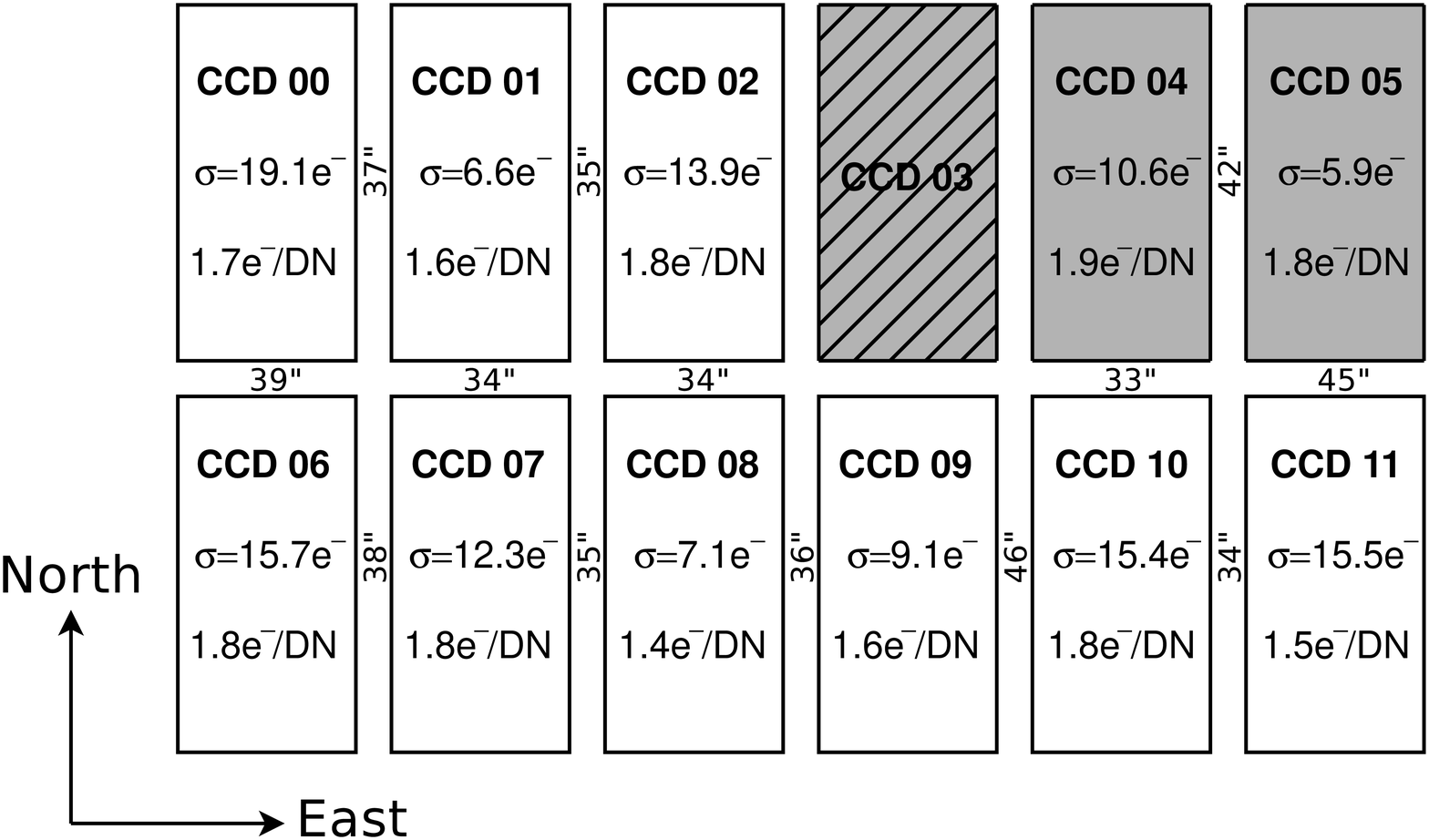}
   }
   \caption{Schematic of the PTF survey camera focal plane (with greatly expanded chip gaps). The HiRho chips (shaded) are high-resistivity silicon with improved red quantum efficiency (Figure \ref{fig:filters}). The read noise and gain is noted for each chip; some chips show relatively high read noise and are being optimized in ongoing work. CCD03 is unresponsive and will be replaced later in the PTF project. Chip gaps are noted in arcseconds, based on on-sky astrometric solutions.}
   \label{fig:ccds_and_gaps}
\end{figure}

\subsection{CCD Array} 

\begin{figure}
  \centering
  \resizebox{\columnwidth}{!}
   {
	\includegraphics[angle=0]{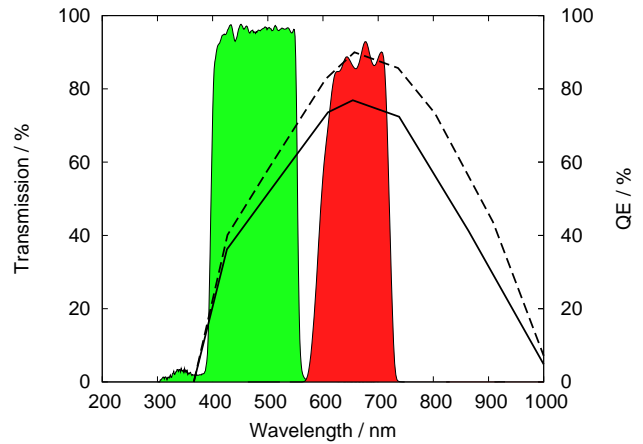}
   }
   \caption{The quantum efficiencies of the nine EPI chips in the CCD array (solid line) and the three HiRho chips (dashed line). Filter transmission curves for the two primary PTF survey filters are also shown (filled curves; green is SDSS-$\rm{g^\prime}$ and red is Mould-R). The QE curves and Mould-R transmission are adapted from \citet{Cuillandre2000}. 
   }
   \label{fig:filters}
\end{figure}

The core of the PTF survey camera is a 12K $\times$ 8K mosaic made up of twelve 2K$\times$4K MIT/LL CCID20 CCDs, arranged in a 6x2 array (Figure \ref{fig:ccds_and_gaps}). Three of the CCDs are high resistivity bulk silicon (HiRho); the rest are standard epitaxial silicon (EPI). The EPI chips reach QEs of approximately 70\% at 650 nm; the HiRho devices have somewhat higher QE, reaching $\approx$90\% at 650 nm (Figure \ref{fig:filters}). The HiRho chips also have lower fringing levels due to their increased thickness. 

Depth of focus, and thus the flatness of the CCD array, is an important issue for the P48's f/2.5 beam. Approximately 92\% of the PTF survey camera's CCD array is within 20 $\mu$m of the reference focal plane position. Taking a detailed error budget into account, including terms for telescope jitter, atmospheric turbulence and the optical quality, we predicted that 89\% of the array would provide images meeting our 2.0\arcsec FWHM image quality budget and that the remainder of the array would be within 0.2\arcsec of the specification. On-sky tests confirm this (\S \ref{sec:image_quality}; the best images so far taken have a median FWHM of 1.5\arcsec across the array). The instrument tip and tilt was adjusted to better than 0.003 degrees tilt across the entire array using 20 $\mu$m shims at the dewar mounting points and on-sky image tilt measurements from FWHM maps. The final full-field-of-view image quality is better than 2.0 arcsec FWHM in median seeing (\S \ref{sec:image_quality}). 

\begin{figure*}
  \centering
  \resizebox{\textwidth}{!}
   {
	\includegraphics[angle=0]{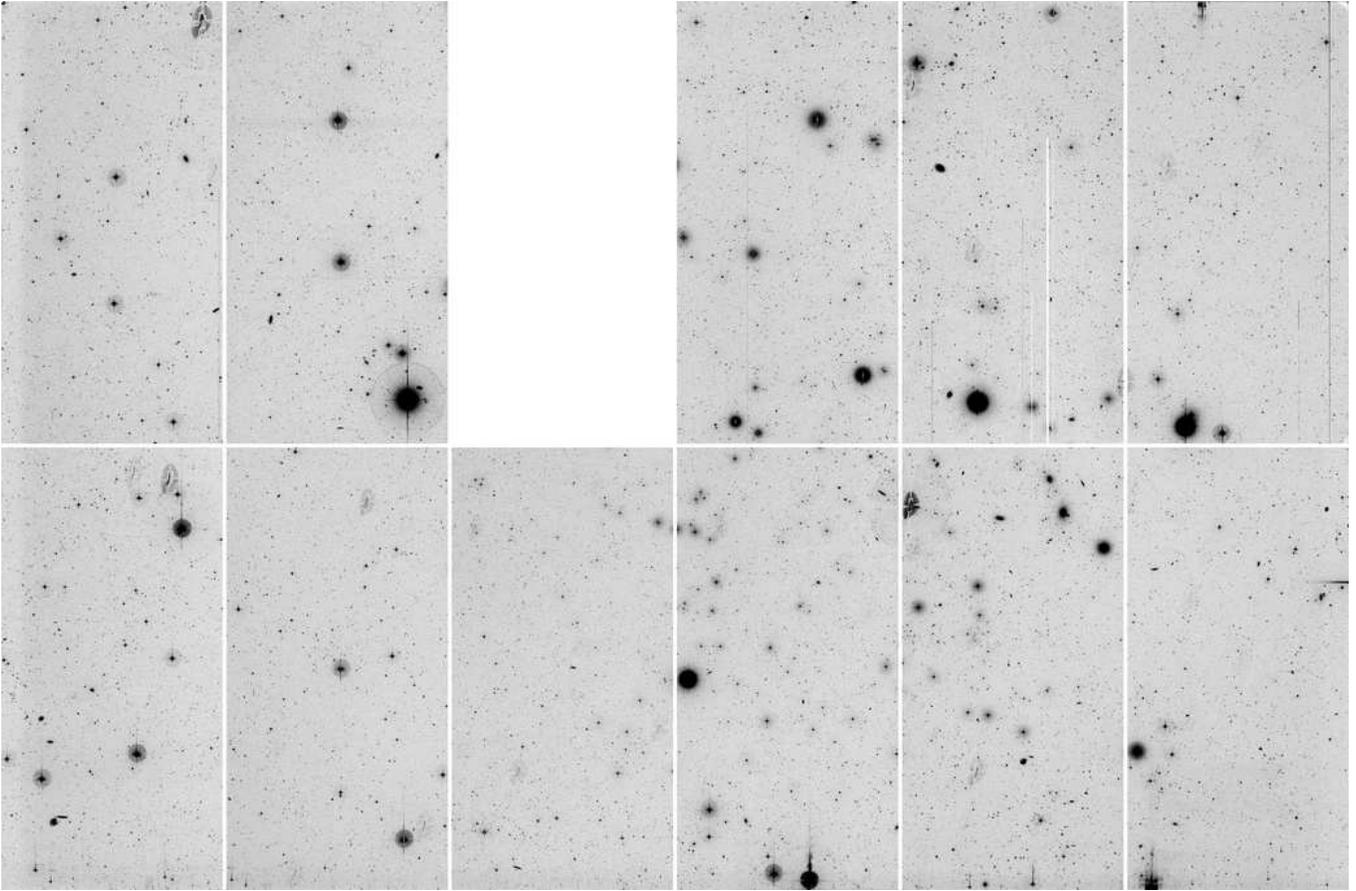}
   }
   \caption{A typical R-band image taken with the PTF survey camera on the Palomar 48" telescope, with the same orientation as Figure \ref{fig:ccds_and_gaps}. The image has been bias-corrected and flat fielded but, to show the cosmetic quality of each chip, no dithering has been applied.}
   \label{fig:typical_image}
\end{figure*}

The CCDs are 3-edge buttable and are separated by gaps of $\approx$500 microns, corresponding to 35 arcseconds (see Figure \ref{fig:ccds_and_gaps} for more details).  The CCD array itself has a low number of bad pixels and offers excellent image quality across the field of view (Figure \ref{fig:typical_image}).

The full CCD array readout time is 30.7 seconds, using independent amplifiers for each chip and two Generation 3 Leach (SDSU) devices reading out 6 CCDs each. This readout time is approximately half that of the original camera electronics and was specified to match the expected telescope slew and settle time for PTF operations. The improved speed was reached at a small cost in readout noise; even so the final readout noise is $<$15$\rm{e^-}$ on all chips, well below the expected sky noise.

During the camera upgrade process CCD03 became unresponsive. After diagnosing the fault as a problem inside the CCD package itself we decided that attempting to repair the CCD would be both an unacceptable risk to the array and an unacceptable delay to the project. We are continuing to evaluate ways to repair or replace the malfunctioning CCD, but for at least the first year of science operations PTF is accepting the 8\% loss of imaging area.

\subsection{Optics}

\begin{figure}[h]
  \centering
  \resizebox{\columnwidth}{!}
   {
	\includegraphics[angle=0]{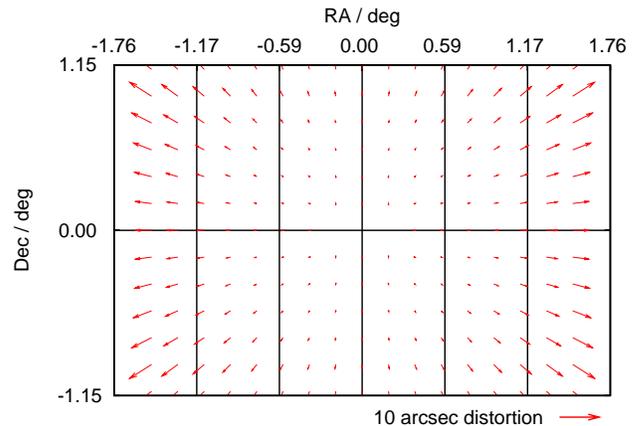}
   }
   \caption{Distortion map across the PTF survey camera focal plane, relative to a flat grid on the sky and based on Zemax modeling of the  optical system. A 20" distortion scale is indicated at the bottom of the figure. The maximum distortion at the corners of the CCD array is 7\arcsec, or 0.11\%.}
   \label{fig:distortion}
\end{figure}

We replaced the original flat dewar window with a field flattening optic designed for P48's curved focal plane. The new optics are designed to provide better than 2\arcsec\, image quality over the entire 3.4-degree span of the image in median seeing, including atmospheric, optical and mechanical tracking error terms. The optical distortion is 0.11\% at the corners of the array relative to a uniform grid, a level comparable to differential atmospheric refraction over our field of view (Figure \ref{fig:distortion} shows the distortion field).

The telescope optics were also optimized: the primary mirror supports were tuned; all optical surfaces were cleaned or refurbished; baffles were improved to control stray light; the telescope tube was made light tight; and internal stabilized calibration sources were added to allow daytime quality control.

\subsection{Cooling system} 
The original liquid nitrogen cooling system was replaced with a closed-cycle cooling system, implemented via a compact cold head located close to the CCDs. A Polycold Compact Cooler (formerly known as CryoTiger) with an expansion head with no moving parts was chosen to minimize vibration. We attached an H-shaped copper head spreader to the cold head to provide a nearly isothermal surface to which to attach flexible copper heat exchange straps leading to the CCD array assembly. A heating system attached to the dewar window actively maintains the CCD array temperature at 175K.

\subsection{Shutter}
The shutter is a commercial unit supplied by Scientific Instrument Technology, and employs dual split blades to avoid beam obscuration by the blades when the shutter is open. The blade transit time is 1.300 seconds, but the blades can be independently controlled to allow shorter exposure times in a slit scanning across the detector. 

 \subsection{Filter changer}
The filter changer assembly is a custom-built mechanism that provides for motorized selection of one of two filters; one filter is always in the field. Filter exchanges are completed in 15 seconds, using a micro-stepper motor. Absolute filter positioning is registered by running the filter frame against a mechanical hard stop. Normal PTF operation is expected to use at most two filters in each night, but the system has been designed for easy manual swapping of filters as necessary.

\subsection{Camera Electronics and Software} \label{sec:camera_soft}
The PTF survey camera array of twelve CCDs is divided into two banks of six CCDs, with each bank handled separately by its own detector controller. The controllers are Generation 3 Leach 
(SDSU) devices comprised of three 2-channel video boards, three clock boards, 
and a timing board. Communication between a controller and its respective host computer is 
conducted via a fiber optic serial link between the timing board and a corresponding PCI card 
in the host computer. The controllers share a common clock so that the exposures and 
readouts of the two controllers are synchronized with each other. The master camera computer also handles all the non-detector camera hardware -- the shutter, filter changer, temperature sensors and calibration LEDs. The camera control software is custom-built for the PTF survey camera, and is based on panVIEW, the Pixel Acquisition Node (PAN) version of ArcVIEW \citep{2002SPIE.4848..508A}.

\subsection{PTF Observatory Control System and Scheduler}

The survey operations are performed robotically by the P48 Observatory Control System (OCS),
written in MATLAB. The OCS is responsible for controlling the camera, filter exchanger,
shutter, telescope, dome, and focuser, on the basis of feedback information from the scheduling system, the camera and telescope, a weather station, and the data quality monitor (\S \ref{sec:dqm}).

The OCS is responsible for sequencing all PTF observations, starting with the bias frames before sunset, through focus images, and finally the science images. The PTF scheduler is responsible for selecting the next target for observation. The PTF schedule is not pre-defined; the next target is selected 30-90\,s before each exposure starts, during the previous exposure.

The criteria for selection of the next target include:
(i) Sun altitude;
(ii) excess in sky brightness due to the Moon, using the algorithm of \citet{1991PASP..103.1033K};
(iii) Moon phase;
(iv) time needed to move telescope to the next target;
(v) time needed to move dome to the next target;
(vi) airmass; and 
(vii) time since the last sequence of observations of this field.
In addition for some programs, such as the 5-day-cadences (\S \ref{sec:5DC}),
the scheduler is required to optimize the field selection so that each field is observed twice during the night for asteroid detection.

These criteria are used to calculate weights for each field, and the scheduler selects the target
with the highest weight for observation. In order to keep the operations of the scheduler
as simple as possible most of the weights follow a step function. An exception is the cadence (time since last observation) weight ($W$) which is of the form:
\begin{equation}
W = 1 - \frac{1}{1+\exp(\{[JD-JD_{l}]-\tau\}/s)}
\end{equation}
where $JD$ is the current Julian day, $JD_{l}$ is the Julian day corresponding to the last successful sequence (pair of images) of observation, $\tau$ is the cadence, and $s$ is a softening parameter
that defines the time scale in which the weight goes near 0 to near 1.

\subsection{Data Quality Monitoring}
\label{sec:dqm}

A software suite running at the mountain, the Data Quality Monitor (DQM), analyses all the recorded images as they are taken. An initial check searches for problems with file corruption, truncation, or missing FITS headers. Next, the image is searched for stellar objects [using SExtractor \citep{bertin96}], and the average FHWM and other image quality parameters are checked for obvious problems. Finally, if it has passed all the checks, the file is sent to the PTF pipelines.

To detect more subtle image quality problems such as increased electronic noise the P48 twice-daily automatically takes test exposures of the telescope tube cover. A band of temperature-stabilized LEDs provides illumination stable to $<$0.1\%. Because the illumination has passed through every optical element of the system, comparing the test exposures to reference exposures provides a simple test if anything in the telescope + camera + electronics systems is not operating as expected. Many problems can be found and corrected before the start of nighttime operations.

Finally, both the LNBL and IPAC pipelines produce image quality and sensitivity measures for all detected fields, providing a long-term monitoring facility to allow us to optimize our observing setup and equipment.

\subsection{On-sky Performance}
\label{sec:image_quality}

The PTF survey camera system achieved first light on 13 Dec 2008. Initial commissioning work was completed in January 2009 and the system has been used for initial science test operations since then (with small gaps for engineering). Here we describe on-sky performance results from the first two months of operation. Since the system continued to be commissioned during this time, the performance is likely to improve as the PTF survey continues.

A total of 91.7\% of the pixels in the full 12K$\times$8K array are light-sensitive and not dark, occulted or otherwise bad (this increases to $>$99\% good pixels if we exclude the unresponsive CCD). Most of the approximately 200 bad columns in the array are concentrated on CCD05 (at one corner of the field). CCD09 offers the best combination of PSF quality and CCD cosmetics. The chips are linear to better than 0.5\% up to 60,000 ADUs (typical gain is 1.6$\rm{e^{-}}$ $\rm{{ADU}^{-1}}$).

\begin{figure}
  \centering
  \resizebox{\columnwidth}{!}
   {
	\includegraphics[angle=0]{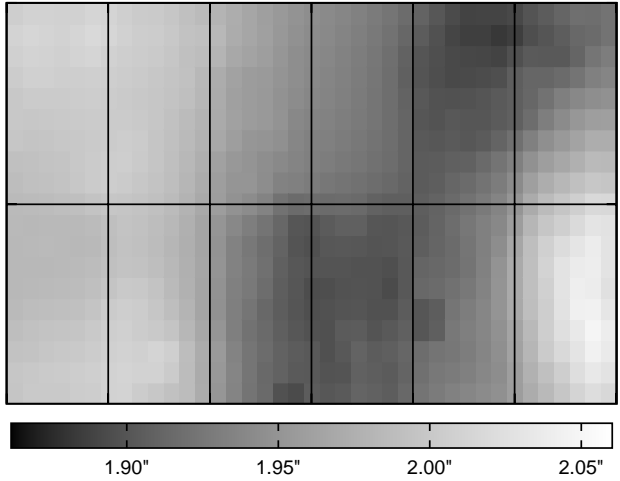}
   }
   \caption{A map of the FWHM variations across the PTF survey camera focal plane array from a typical R-band observation (with the same orientation as Figure \ref{fig:ccds_and_gaps}). Image FWHM measurements are interpolated over the unresponsive CCD for ease of interpretation. The residual FWHM variations are on the order of 0.2\arcsec and are due to the 20$\mu$m flatness of the CCD array and slightly decreasing optical performance at the edge of the field of view.}
   \label{fig:fwhm_map}
\end{figure}

\begin{figure}
\begin{center}
 
   	\subfigure{\includegraphics[width=0.325\columnwidth]{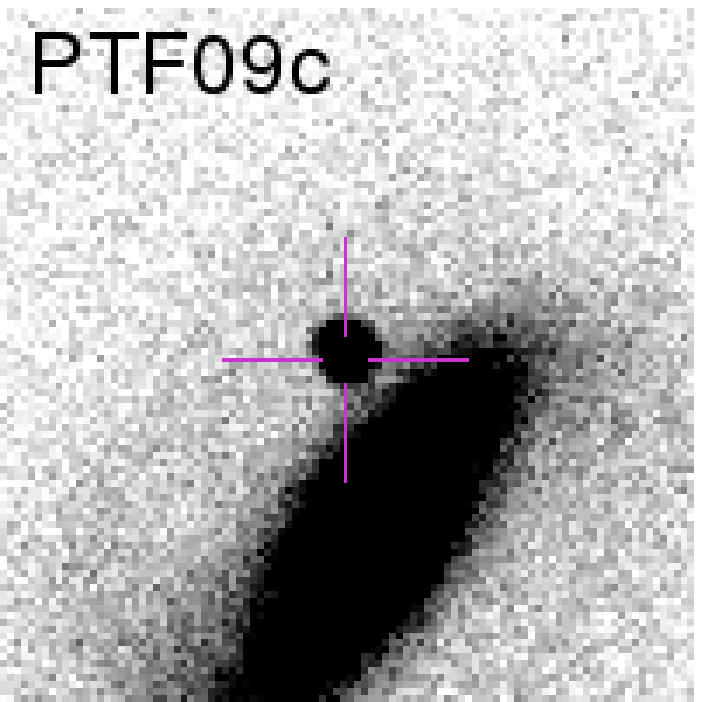}}
   	\subfigure{\includegraphics[width=0.325\columnwidth]{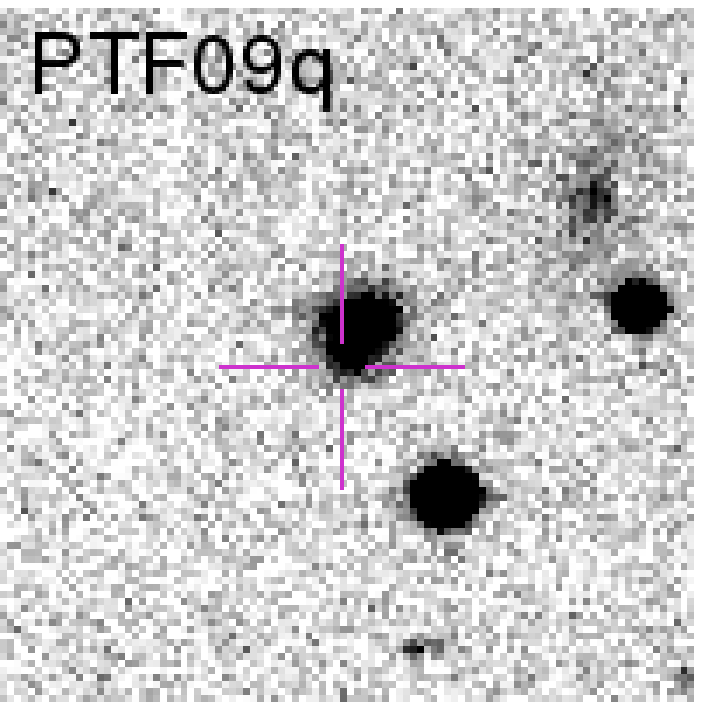}}
   	\subfigure{\includegraphics[width=0.325\columnwidth]{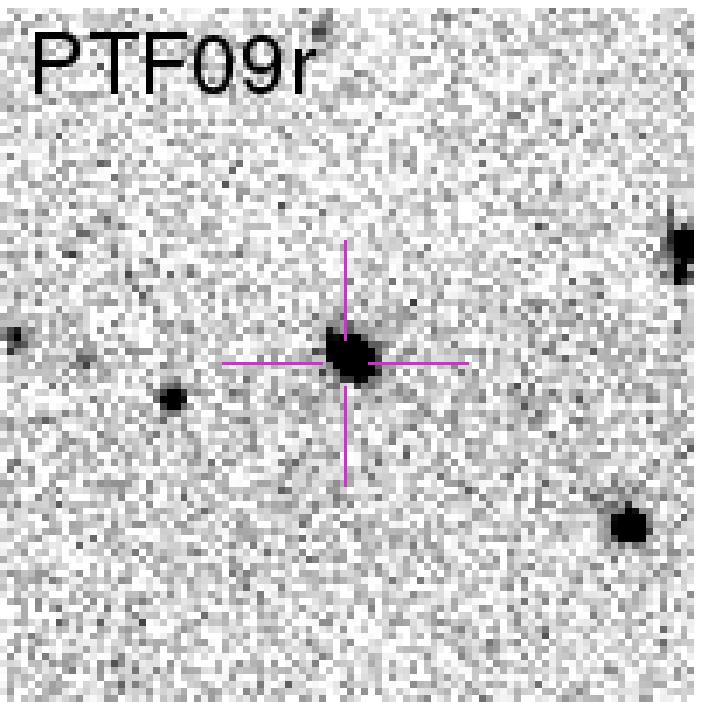}}
   	\subfigure{\includegraphics[width=0.325\columnwidth]{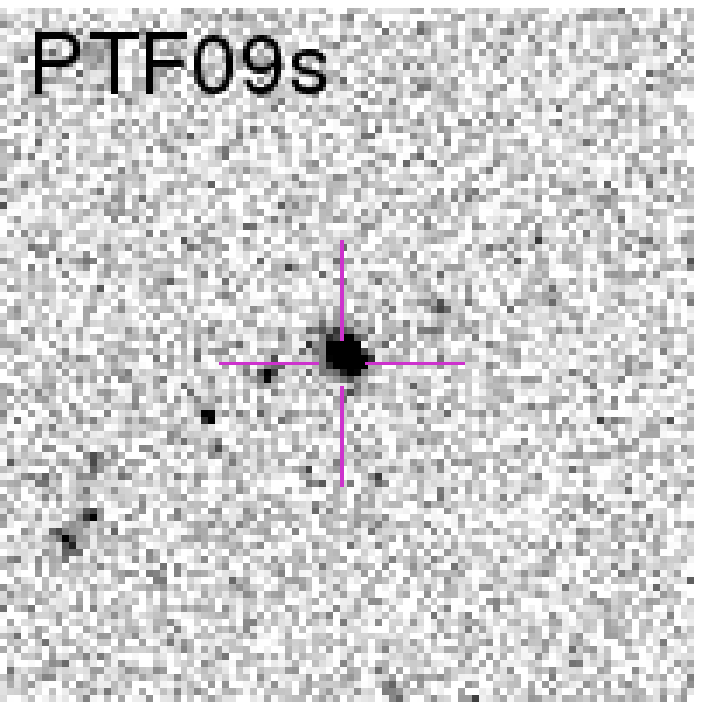}}
   	\subfigure{\includegraphics[width=0.325\columnwidth]{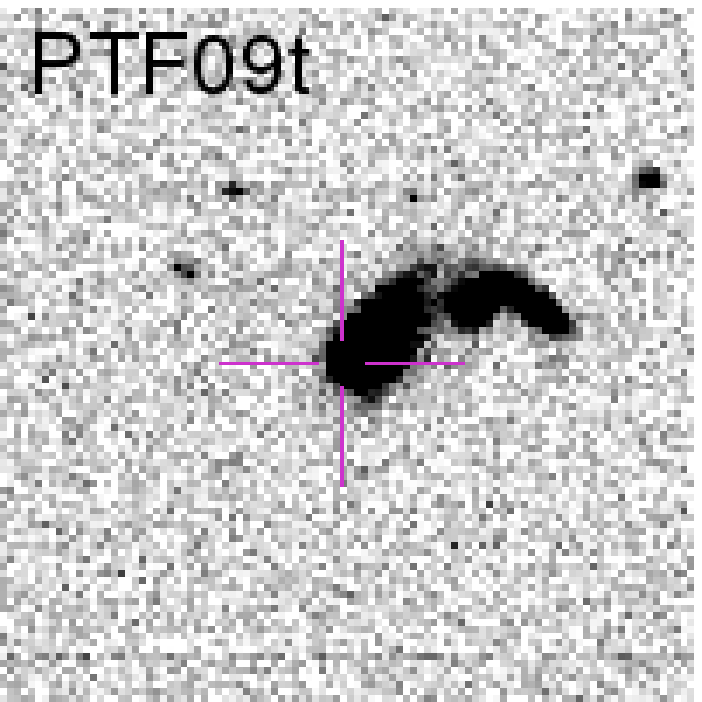}}
   	\subfigure{\includegraphics[width=0.325\columnwidth]{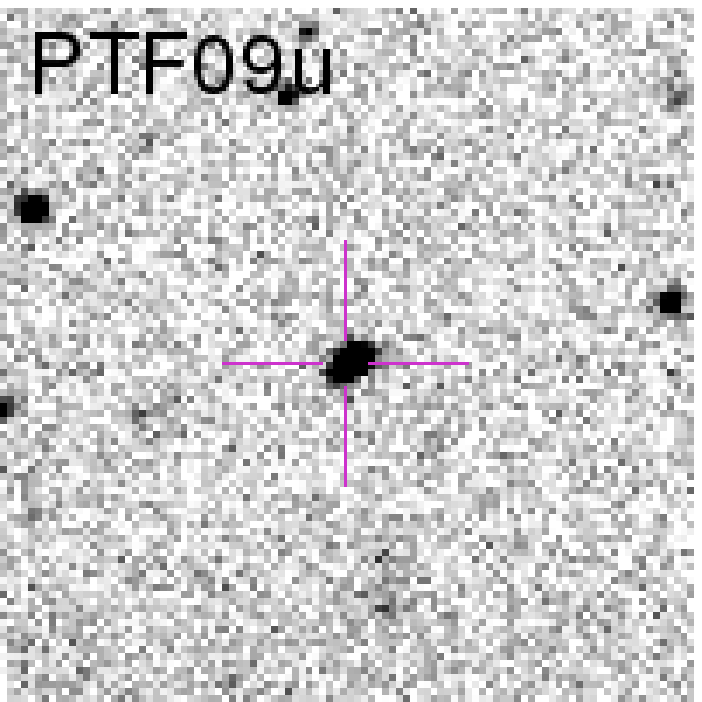}}
   	\subfigure{\includegraphics[width=0.325\columnwidth]{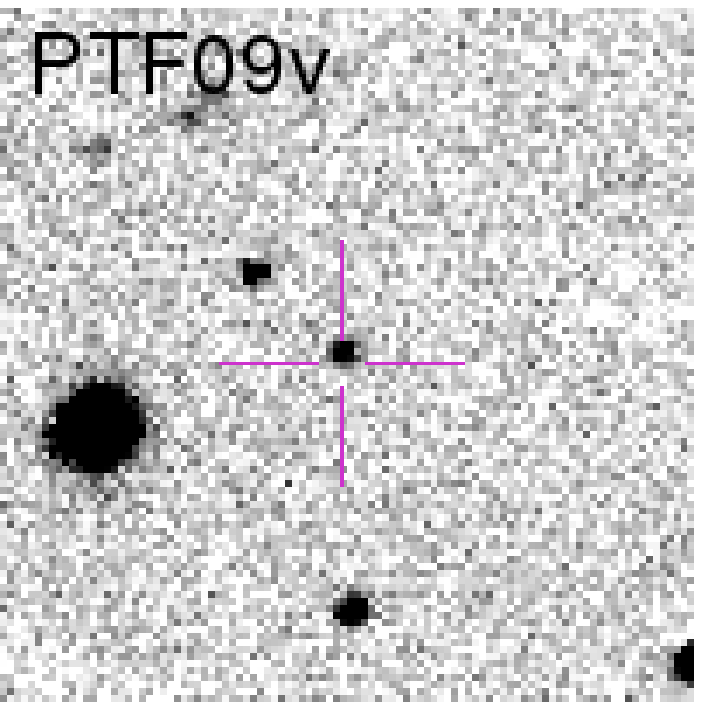}}
   	\subfigure{\includegraphics[width=0.325\columnwidth]{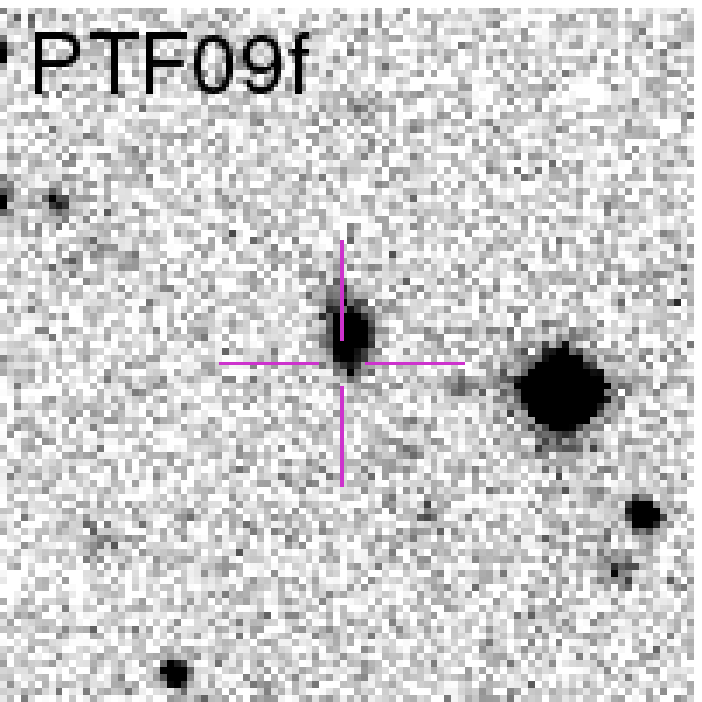}}
   	\subfigure{\includegraphics[width=0.325\columnwidth]{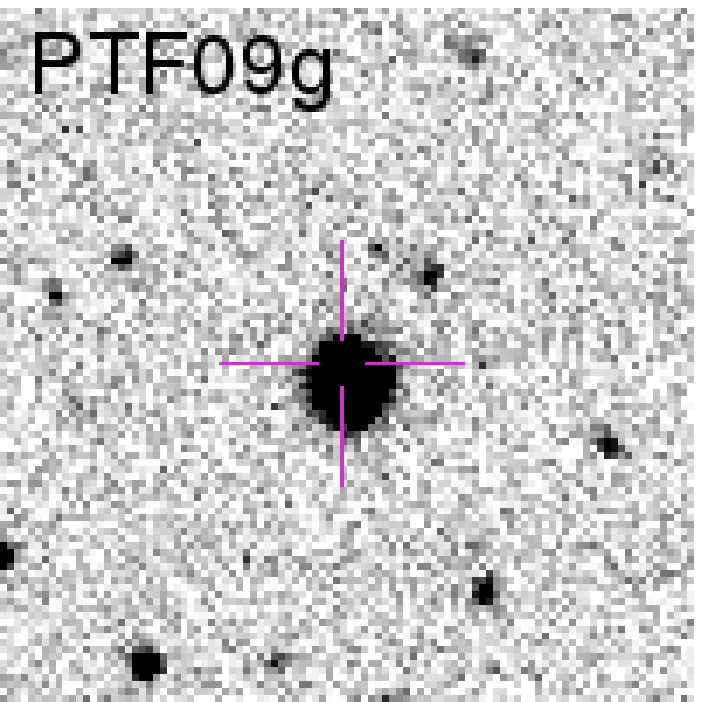}}
\end{center}
   \caption{100\arcsec cutouts from R-band PTF survey camera images, from the first set of PTF optical transient detections. Crosshairs show the transient detection position, most inside galaxies. The properties of the detected transients are given in \S\ref{sec:sn}, with the exception of PTF-OT3, which is a re-discovery of SN2009an. The images are distributed widely across the focal plane.}
   \label{fig:example_images}
\end{figure}

During the design of the PTF survey camera we specified a FWHM of 2.0\arcsec\, during median Palomar conditions (1.1\arcsec\, seeing). This requirement was based on a number of factors -- the minimum PSF size that is reasonably sampled by our 1\arcsec/pixel plate scale; what could be reasonably achieved with the P48 optical setup; optimization of the survey limiting depth; and the requirement to detect supernovae in the cores of bright nearby galaxies. Analysis of the first two months of PTF on-sky data shows that we have met or exceeded this requirement. Typical images taken in good seeing show both sub-2.0\arcsec\, FWHMs and very small variations in image quality across the entire field of view (Figures \ref{fig:fwhm_map} \& \ref{fig:example_images}). 

\begin{figure}
  \centering
  \resizebox{\columnwidth}{!}
   {
	\includegraphics[angle=0]{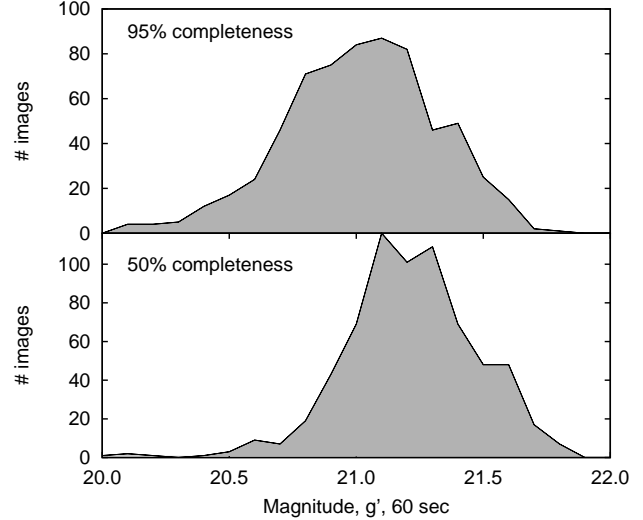}
   }
   \caption{Completeness limits from a typical $\rm{g^\prime}$-band, photometric, PTF survey night. Each curve represents the minimum detectable  magnitude at either 50\% or 95\% completeness; each exposure of each chip is treated as an individual image. The spread in values arises from variations due to seeing, the imaging quality of each chip and the airmass of the observations, and thus represents typical performance for the actual PTF observation strategy.}
   \label{fig:lim_mags}
\end{figure}

The survey depth is correspondingly encouraging (as shown in Figure \ref{fig:lim_mags}). Based on all the data taken during the last two months of commissioning, the system provides median 5$\sigma$, 60-sec limiting dark-time magnitudes of $\rm{m_{g^\prime}}$=21.3 ($\rm{25^{th}}$ percentile $\rm{m_{g^\prime}}$=21.7; $\rm{75^{th}}$ percentile $\rm{m_{g^\prime}}$=21.0) and $\rm{m_{R}}$=20.6 ($\rm{25^{th}}$ percentile $\rm{m_{R}}$=20.8; $\rm{75^{th}}$ percentile $\rm{m_{R}}$=20.3).

The entire PTF survey system (the P48 telescope, PTF survey camera, OCS and the DQM) has a current uptime of $\approx$90\%, excluding losses due to weather. As initial science operations proceed, many issues are being addressed and the uptime is continually improving. The survey efficiency is currently approximately 50\% open-shutter time, compared to a goal (and theoretical maximum) of 66\%. This is due to the cumulative effect of many small inefficiencies in the system operation and is continually improving.

\section{The PTF Observing Strategy}
\label{sec:strategy}
PTF will pursue four distinct experiments optimized for different times of the year. Each project is designed to use P48 only as the survey telescope while other PTF resources are reserved for followup observations. We here describe the observing strategy for each of the major PTF experiments; \citet{Rau2009} give a detailed discussion of the science planned for each project. The experiments are summarized in Table \ref{tab:strategy_sum}.

\begin{table}
\caption{The PTF experiments planned for the first year of operations.}
\centering
\label{tab:strategy_sum}
\begin{tabular}{lllll}
Experiment & \% of total & Cadences & Filter & Months\\
\hline
5-day cadence (5DC) & 41 & 5 d & $R$ & Mar-Oct \\
Dynamic cadence (DyC) & 40 & 1 min - 5 d & $\rm{g^\prime}$, $R$ & all year\\
Orion & 11 & 1 min & $R$ & Nov-Jan\\
H$\alpha$ & 8 & -- & H$\alpha$ & all year\\
\end{tabular}
\end{table}

\subsection{Observing Setup \& Filter Selection}

The three major PTF experiments, together accounting for 92\% of the project's P48 time, are based on two filters: SDSS $\rm{g^\prime}$ and Mould-R (Figure \ref{fig:filters}). These filters were chosen to optimize the survey detection limits in full moon (R-band) and dark sky ($\rm{g^\prime}$-band) conditions for a variety of different transient classes (see \citealt{Rau2009}). During dark-sky conditions, the reduced quantum efficiency in the $\rm{g^\prime}$ filter is more than compensated for by the darker sky and the blue colors of many of our targeted transients. Mould-R is very similar to SDSS $\rm{r^\prime}$ and allowed us to reuse an existing CFH12K filter; it is the standard filter for programs that are expected to spend at least part of their time operating while significant moonlight is present. Because no followup is done with P48, we elected to use standard passbands for the survey to allow a meaningful comparison of PTF detection magnitudes to followup telescope results.

The initial standard exposure time for the 5DC and DyC surveys is 60 seconds, based on the readout time of the camera, expected system efficiency, and the expected sky noise and dark current in the exposures. In later stages of the PTF project this exposure time will be optimized for the different programs expected to be pursued.

\subsection{5 Day Cadence}
\label{sec:5DC}
The 5-day cadence experiment (5DC) will run yearly from March 2nd until October 30th using 65\% of the time in that period. The main goal of this experiment is to construct large samples of type-Ia 
SNe and core-collapse SNe. It will also study AGNs, Quasars, Blazars, CVs, extragalactic novae, luminous red novae, and tip of the AGB variability.  The experiment will observe a footprint of approximately 8000 square degrees, with $|b| > 30^{\circ}$, ecliptic latitude $|\beta| > 10^{\circ}$, and with a median cadence of 5 days. At a given time, the active area for the 5DC search (including all overheads) will be 2700 square degrees. The 5DC observations will be conducted in almost all lunar phases in R-band. In each epoch we will obtain two 60 s exposures separated by 60 min. The two images will be used to remove cosmetic artifacts, cosmic rays and to find solar system objects.

\subsection{Dynamic Cadence}
The dynamic cadence experiment (DyC) is designed to explore transient phenomena on time scales shorter than $\sim$5 days and longer than one minute. The experiment will explore unknown territories in the transients phase space -- mainly fast transients with time scales of less than a day. Therefore, the dynamic cadence will be an evolving experiment. Every six months the PTF collaboration will review the results of the dynamic cadence experiment and decide how to continue. 

Motivated by a search for short-lived transients in the brightness gap between novae and supernovae($-$10$<$M$<-$16), the DyC experiment is targeted at fields that are luminosity concentrations in the local universe. Since these transients are expected to be fainter than M$\sim-16$, the search volume is limited to DM (distance modulus) $<$ 36.5 -- i.e. less than 200\,Mpc.

The total luminosity of nearby galaxies in the PTF FoV was computed on a fine grid
of possible pointings (limited by declination $>-$30). This was done seven times with seven different maximum distances (in steps of 0.5 between 30.5 $\leq$ DM $\leq$ 36.5). We assigned a weight for each pointing based on the fraction of maximum luminosity in that list. A total weight for each pointing was computed by averaging the individual weights. Finally, pointings with the highest weights were chosen for PTF. This list naturally includes the nearest, brightest galaxies (e.g. M31, M51, M81) and nearest
galaxy clusters (e.g. Virgo, Perseus, Coma), but also covers the other accessible nearby mass concentrations from physical and chance galaxy alignments.

This experiment will detect RR Lyr stars in the Galactic halo, CVs, AM CVn stars, flare stars, SNe, eclipsing binaries, and solar system objects. Moreover, this experiment will put strong constraints on the existence of any optical counterparts that may be associated with radio transients.

\subsection{Orion Field}
The Orion project has been assigned 40 consecutive nights per year for
three years to perform intensive time-series observations of a single field in the Orion star forming region, with the aim of detecting close-in Jupiter-sized planets transiting young stars.  The observations will be acquired in R-band with no dithering in a continuous sequence of 30-second images, with the goal of producing high precision (better than 1\%) differential photometry. The field will be chosen to represent a stellar age range where protostellar discs should be on the point of dissipation (5-10 Myr) -- giving important clues about planet formation and migration -- and to optimize for source density and avoid regions where the stellar variability in the ensemble is severe enough to prohibit precise differential photometry. The IPAC PTF pipeline will be used for standard image reduction, source identification, and astrometry, but a dedicated photometric pipeline is being developed to produce the high-precision differential photometry.

\subsection{Deep H$\alpha$ Sky Survey}
For three nights each lunation during full Moon we will carry out a $3\pi$\,sr sky survey in H$\alpha$. The survey area will include all the sky with $\delta>-25^{\circ}$.
Each field will be observed twice in four narrow-band (approximately 10 nm width) filters around the H$\alpha$ line covering the redshift range of about zero to 0.05, corresponding to distances of 0 to 200\,Mpc. Narrow-band filters that give a consistent bandpass across the entire field of view of the camera are difficult to produce, especially in the fast beam of the P48. However, a description of usable H$\alpha$ filters for a Schmidt telescope with the same optical design is given in \citet{1998PASA...15...33P}, and our industrial partners are confident that these filters can be constructed.

We estimate that the H$\alpha$ survey sensitivity limit will be $\approx$0.6R in a 2.5 arcsec aperture\footnote{R is a unit of surface brightness commonly used in aeronomy. One R (one Rayleigh) is $106/(4\pi)$ photons per square centimeter per steradian per second.
For the H$\alpha$ line the intensity in cgs units is $2.41\times10^{-7}\,{\rm erg\ cm^{-2}\ s^{-1}\ sr^{-1}}$.}; a detailed comparison of this sensitivity to other H$\alpha$ surveys can be found in \citet{Rau2009}. We are currently designing a scheme that will allow us to photometrically calibrate these data to about $10\%$ accuracy.

\section{Data Reduction and Transient Detection}
\label{sec:pipelines}

All PTF data taken at P48 is automatically routed to two pipelines: a realtime transient detection pipeline, optimized for rapid detection of interesting objects, and a longer-term archival pipeline optimized for high precision and easy searching of the results.

\subsection{Realtime Transient Detection}
\label{sec:realtime_detection}

The PTF quick-reduction and subtraction pipeline was designed from the
start to take full advantage of the parallel high performance
computing facilities at the National Energy Research Scientific
Computing Center (NERSC, http://www.nersc.gov). The major
improvement to the hardware used in this pipeline over previous
versions designed by the Supernovae Cosmology Project \citep{1999ApJ...517..565P} and Nearby Supernovae Factory \citep{2002SPIE.4836...61A} at LBNL is the use of the NERSC Global
Filesystem (NGF).  NGF is a 300TB shared filesystem with over 1 GB/sec
bandwidth for streaming I/O which can be seen by all of the high
performance computers. On the software front, the improvements include
a tight coupling of a PostgreSQL database involved in tracking every facet
of the image processing, reference building, image subtraction and
candidate detection along with a complete re-write of the processing
and subtraction codes, described below.

Incoming P48 images are immediately backed up on the High
Performance Storage System, a 3 PB tape archive system at
NERSC. Processing begins by splitting the packed images apart by chip,
applying crosstalk corrections, and performing standard
bias/overscan subtraction and flatfielding. After this point, all
operations are performed on a chip-by-chip basis in
parallel. Catalogs of sources are created for each image via the
Terapix SExtractor code \citep{bertin96}, which then is fed to the astrometry.net (http://astrometry.net)
code to perform an astrometric solution. At this point a comparison to
the USNO catalog is made to determine the zeropoint and 3-$\sigma$
limiting magnitude of the image along with a calculation of the
seeing. The image is then loaded into the processed image database,
storing all relevant information from the FITS headers.

\begin{figure}
\begin{center}
 
   	\subfigure{\includegraphics[width=0.32\columnwidth]{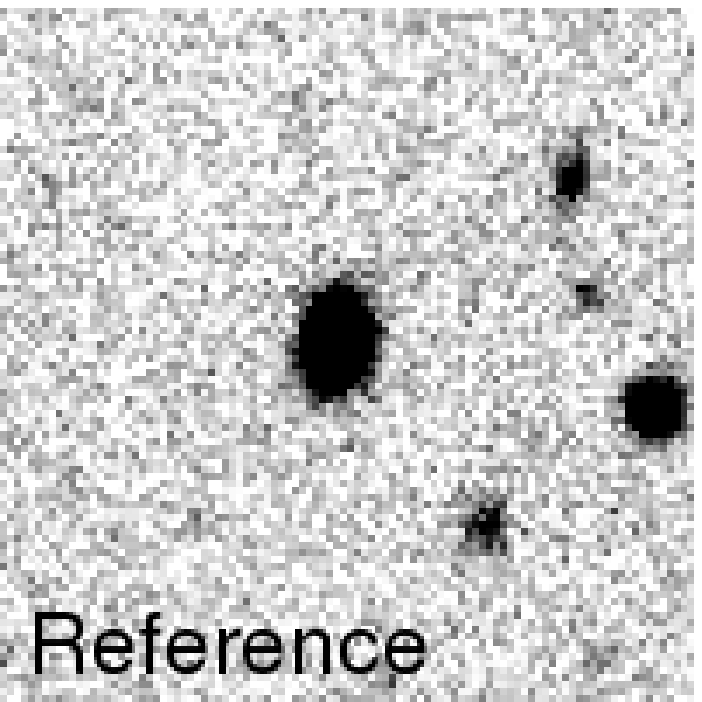}}
   	\subfigure{\includegraphics[width=0.32\columnwidth]{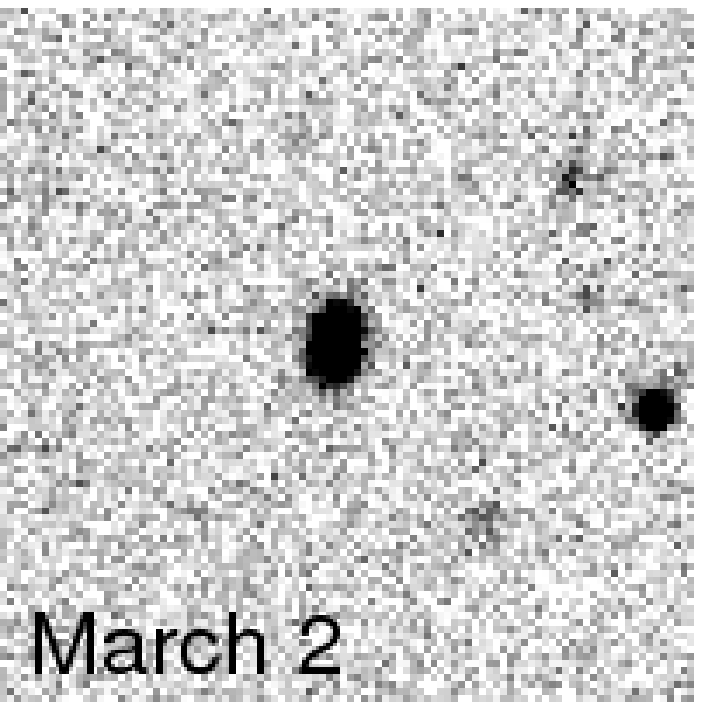}}
   	\subfigure{\includegraphics[width=0.32\columnwidth]{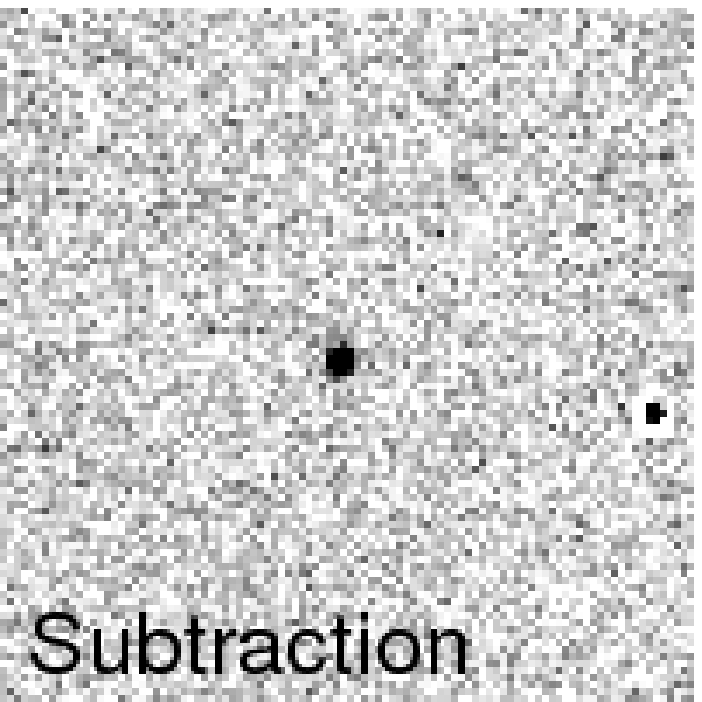}}
	
   	\subfigure{\includegraphics[width=0.32\columnwidth]{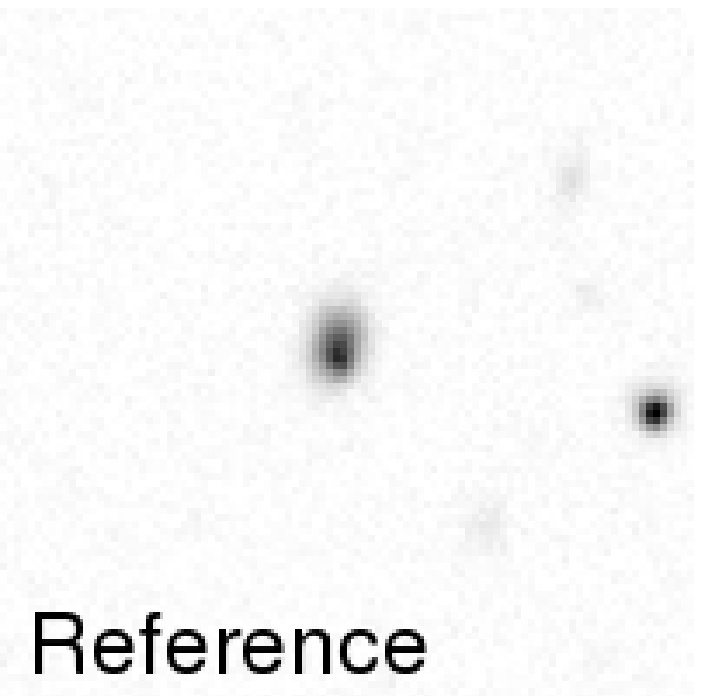}}
   	\subfigure{\includegraphics[width=0.32\columnwidth]{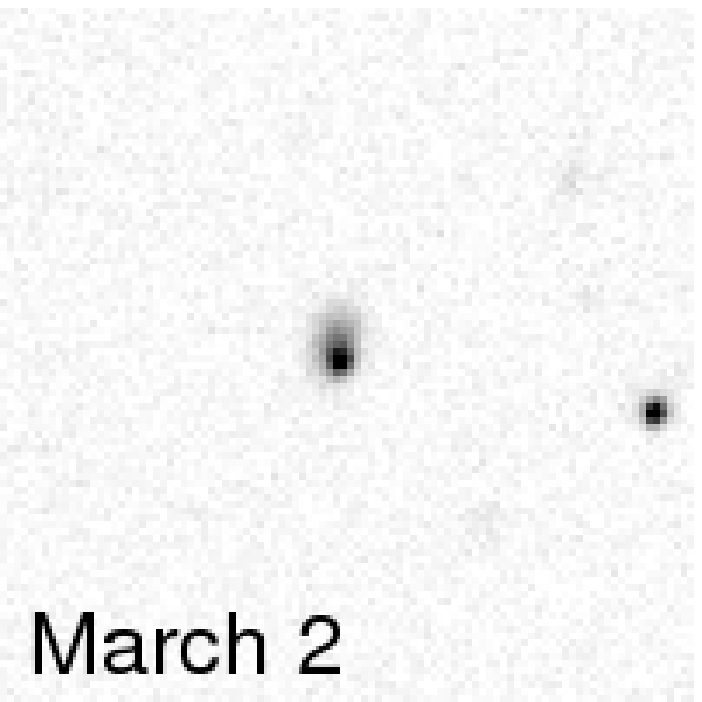}}
   	\subfigure{\includegraphics[width=0.32\columnwidth]{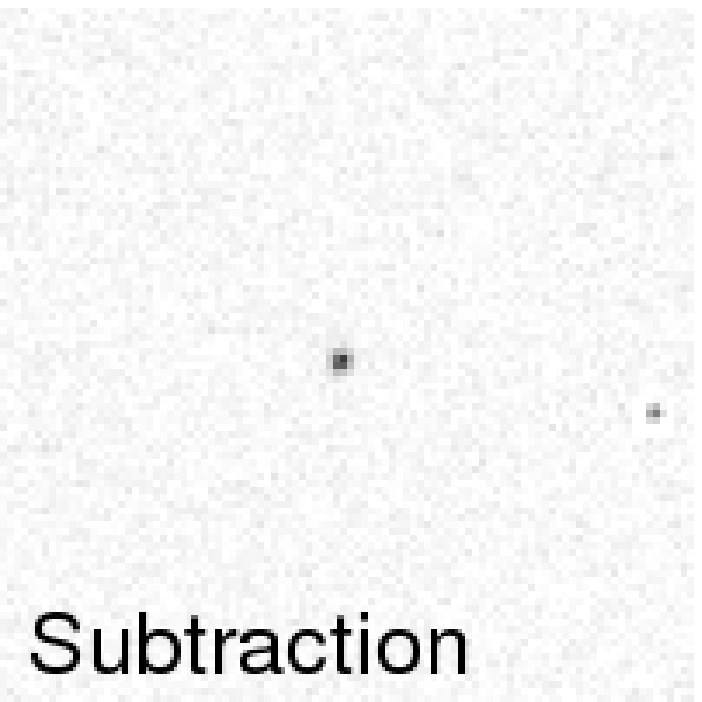}}
\end{center}
   \caption{The detection images of SN 2009av, PTF's first confirmed transient detection. The top and bottom row show the same images at different stretches.}
   \label{fig:sn_img}
\end{figure}

After several images are obtained for a given PTF pointing, we are
able to create a reference image. The reference images are created via
the Terapix software Scamp and Swarp \citep{bertin2006} after querying the processed
image database in order to obtain the highest quality input images
for a given pointing/chip combination. The references are assigned
version numbers and (along with their relevant information) are stored
in the reference database. 

Subsequent new images are processed as above, and if a reference image
exists for them, a subtraction is performed. First, the new image is
astrometrically aligned to the reference by utilizing Scamp and the
new and reference catalogs. The reference is then Swarped to the size
and scale of the new image and a subtraction is performed using HOTPAnTS\footnote{\tt http://www.astro.washington.edu/users/becker/hotpants.html}. The subtraction is performed in both the positive and
negative direction (reference minus new and new minus
reference) to detect both positive and negative flux changes. Candidate transient sources are detected via SExtractor and along with
other relevant parameters (location with-respect-to bad pixels, etc.) 
are stored in a candidate and subtraction database. The subtraction image that lead to the first PTF supernova discovery is shown in Figure \ref{fig:sn_img}.

In addition, PTF  takes advantage of DeepSky
(http://www.deepskyproject.org). DeepSky was started in response to
the needs of several astrophysics projects hosted at NERSC. It is a
repository of digital images taken with the point-and-stare observations
by the Palomar-QUEST Consortium and the Near Earth Asteroid Team \citep{Djorgovski_2008}. This data spans nine years and more than 15,000 square
degrees, with 20-200 pointings on a particular part of the sky and
cadences from minutes to years. For a large fraction of the survey
DeepSky achieves depths of $\rm{m_R}$ $>$ 23 magnitude. In total there are more
than 11 million images in DeepSky. This historical data set
compliments the Palomar Transient Factory survey by identifying known
variable stars, AGN, and the detection of low surface brightness host
galaxies of supernovae.

During commissioning human scanners were utilized to reject "junk" transient detections via a web-interface. The subtraction system and rejection algorithms are currently being improved with an aim of full automation by making use of machine learning techniques. The results of human scanners are being used to train an automatic classifier; a more detailed account of this classification technique is in preparation.

Any statistical inference of PTF transient rates will require a careful measurement of the survey detection efficiency. We have constructed a pipeline that produces realistic fake variable objects and adds them into PTF images, on which we apply the standard PTF search algorithms to determine the detection efficiency for any specific data subset. We currently model three generic classes of transients: variable stars, SN/AGN-like sources (i.e., point sources in host galaxies), and ``rogue" transients (point sources on random positions in the image). The transients are directly injected into PTF images using PSFs modeled from nearby stars in the fields.

\subsection{Realtime Automated Transient Classification}
The Transients Classification Pipeline (TCP; \citealt{2008AN....329..284B,2008ASPC..394..609S,2008AIPC.1000..635S}) is a parallelized, Python-based framework created to identify and classify transient sources in the realtime PTF differencing pipeline (Section \ref{sec:realtime_detection}).  The TCP polls the candidate database from that pipeline and retrieves all available metadata about recently extracted sources.  Using the locations and uncertainties of the transient candidate objects, the TCP either associates an object with existing known sources in the TCP database or, after passing several filters to exclude non-stellar (e.g. known minor planets) and non-astrophysical events, generates a new source. The Bayesian framework for event clustering is given in \citet{2008AN....329..284B}. Another classification engine will also be used on PTF transients \citep{Mahabal2008, Mahabal2008a}.

Once a transient source has been identified, the TCP generates ``features'' which map contextual and time-domain properties to a large dimensional real-number space.   After generating a set of features for a transient source candidate, the TCP then applies several science classification tools to determine the most likely science class of that source.  For rapid-response transient science, a subset of features --- such as those related to rise times and the distance to nearby galaxies --- are most useful. As light curves are better sampled and colors are obtained  more features are used in the classification. The resulting science class probabilities are stored in a database for further data mining applications.  Sources with high probabilities of belonging to a science class of interest to the PTF group will be broadcast to the PTF's ``Followup Marshal'' (Section \ref{sec:followup}) for scheduling of followup observations.  As more observations are made for a known source, the TCP will autonomously regenerate features and potentially reclassify that source.

\subsection{\label{idp} The IPAC PTF Image Processing Pipeline and Data Archive}

While the transient pipeline will be optimized for rapid turn-around,
the IPAC pipeline will be
geared towards providing the best possible photometry consistent with
hardware capabilities and data rate. To enable the generation of
optimal calibration files (primarily flat fields), the IPAC pipeline
will commence image processing only after all of a given night's
observations have been received. Additional source association
operations, database updates, and data transfer to the Infrared
Science Archive (IRSA (http://irsa.ipac.caltech.edu) will be carried
out over a period of several days. Processed images and incremental
source lists will be available through IRSA within 2 days from the
time the observations were taken.

After a given night's observations have been completed and received at
IPAC (as indicated by receipt of end-of-night image manifest), the
image processing pipeline is initiated. Parallel processing on 11
multi-processor Linux machines is employed to simultaneously process
the PTF mosaic's 11 working detectors and meet the data rate requirement.  Key pipeline-processes are multi-threaded to take full advantage of the 8-core CPU in each of the Linux machines.

The pipeline initially generates night-specific bias frames, and trims
and bias-subtracts all images in the usual manner. Nightly flat fields
are generated for each detector by source-masking and median-combining
all of the night's sky observations. This is an involved and memory
intensive process and, given the number of exposures involved, takes
3-4 hours with the processing hardware currently in place.  On
completion, these flat fields are used to process the entire night's
science observations.  Raw and processed images as well as all
ancillary and calibration frames are subsequently passed to IRSA for
archiving and retrieval on demand.

Source detection is carried out using SExtractor \citep{bertin96}.
SExtractor is applied to each image a number of times in the pipeline
to mask sources prior to flat field generation, measure the effective
seeing, generate pixel weights and masks, and finally to generate
photometry for all detected sources. Pixel masks are generated to flag
bad rows and columns, cosmic rays, saturated pixels, and charge
bleeding. The final application of SExtractor is carried out after
multiplying the processed image by a pixel area map to preserve point
source photometry.

For each image, the SExtractor output list is then astrometrically calibrated using the Terapix Scamp software
\citep{bertin2006}.  Photometric zeropoints are determined by
calibrating against stars in the Sloan Digital Sky Survey
\citep{york2000}, taking color terms into account. With appropriate airmass corrections, these
zeropoints are then be applied to all sources outside the SDSS
footprint. Because the PTF survey camera's CCD conserve signal when stars are bloomed due to saturation, absolute photometric calibration may also be carried out using bright ($\rm{m_V}$$\sim$11) saturated Tycho stars for which the $griz$ magnitudes are known \citep{Ofek2008}. The density of such stars in the PTF survey camera field of view is
high enough and we estimate it will be possible to achieve
accuracies of better than 5\% in both the $\rm{g^\prime}$ and R bands in non-SDSS fields. 

Currently in development is a source association pipeline that will match, on a nightly basis, each detected source with all previous detections of that source. Appropriate pointers will be incorporated into the IRSA database, enabling users to query the photometric time history of every object ever detected by the PTF."

\section{Following-up Detected Transients}
\label{sec:followup}
The automated Palomar 60-inch is the primary workhorse for photometric
follow-up of PTF transients.  Further photometric and spectroscopic characterization is performed on a worldwide network of other telescopes, including the Palomar Hale Telescope, the Keck telescopes, PARITEL, and the LCOGT telescope network.

\subsection{Photometric Followup with P60}
The automated P60 observation queue \citep{Cenko2006} is sent all candidates flagged as
bona fide astrophysical transients (i.e. a new point source detected at least twice in P48 imaging and which either lacks a counterpart or has a galaxy host in the reference imaging). A 120\,s snapshot in g$^{\prime}, $r$^{\prime}, $i$^{\prime}$,
z$^{\prime}$ is obtained for quick classification.  On a typical
night with median weather conditions, this reaches a limiting
magnitude of $\rm{m_g^{\prime}}\approx$21.4, $\rm{m_r^{\prime}}\approx$21.2,
$\rm{m_i^{\prime}}\approx$20.9 and $\rm{m_z^{\prime}}\approx$19.8. Our efficiency
currently allows twenty two-minute exposures per hour.

As soon as the 60-inch data is obtained, an automated pipeline
detrends the data and solves for an astrometric and zeropoint solution.
Next, if the location is in SDSS, the pipeline downloads the corresponding SDSS image (mosaiced to fit the P60 FoV) to serve as a template. A convolution kernel is measured and image subtraction performed. On the
subtracted image, PSF photometry is used to measure the magnitude at the
source location and the result is sent to the PTF database. If the source is not detected, an upper limit is returned. Currently, if the location is outside SDSS, direct photometry is performed. We note that host galaxy light can be a significant contaminant which is not properly corrected for
in case of direct photometry. Hence, even for fields outside SDSS, we are investigating other options for template imaging for image subtraction.

The 60-inch snapshot observation provides independent confirmation that the
candidate is a real transient (and not a subtraction artifact), a color for quick classification (for example, $\rm{m_g-m_r}\approx 0$ suggests that it is a SN\,Ia), and a second-epoch magnitude to assess photometric evolution. All this information helps direct further spectroscopic and multi-band photometric follow-up with P60 and other telescopes.

\subsection{The Followup Marshal}

We have designed a centralized follow-up Marshal to ensure that our discoveries are efficiently distributed to
our follow-up resources and that the temporal coverage of a given
target meets the requirements of the appropriate science program. The design of this system is complete, but the 
automated followup program has not yet begun. A later paper will detail the performance
of this novel system.

\begin{figure*}
  \centering
  \resizebox{0.9\textwidth}{!}
   {
	\includegraphics[angle=0]{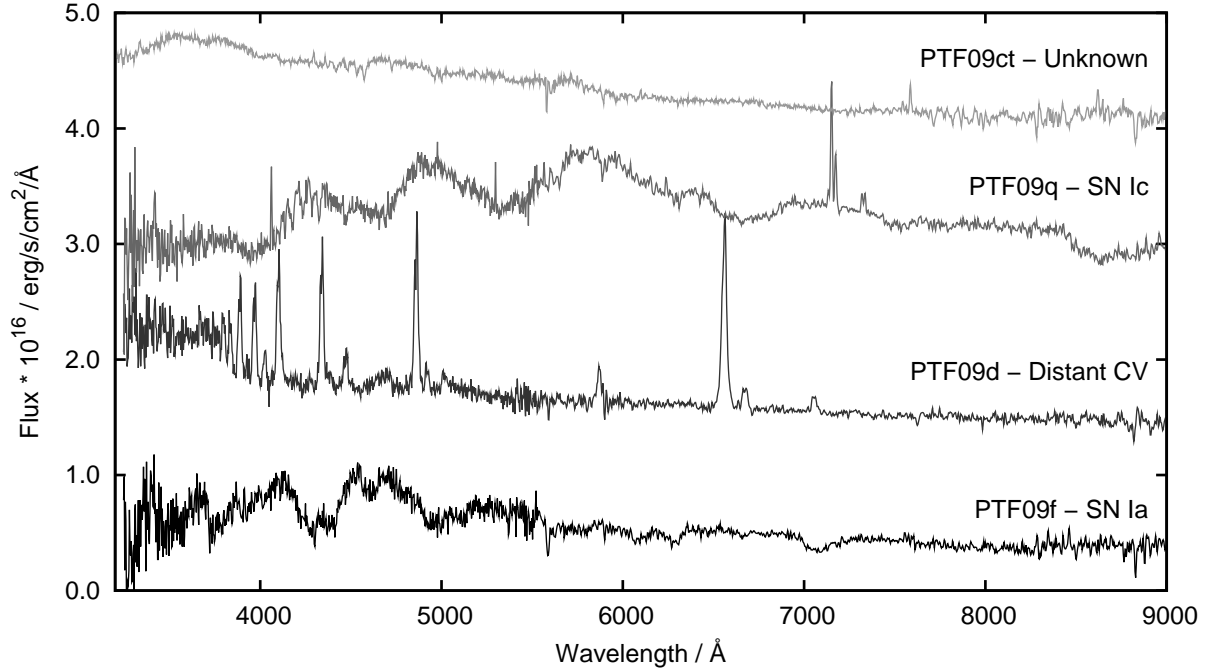}
   }
   \caption{Keck LRIS spectra of several different types of optical transients discovered and classified by PTF. Spectra are not redshift-corrected, and are vertically offset for clarity. Details of the detected transients can be found in Table \ref{tab:march_transients}. }
   \label{fig:sn_spect}
\end{figure*} 

The first duty of the follow-up Marshal will be to monitor the output
of the transient classifier to identify any new sources detected by
P48 that warrant follow-up observations. For this purpose, the Marshal
will maintain a list of target descriptions (in terms of the
classifier output) for each science program to define trigger
conditions for same night and long term monitoring. For example, a
supernova program may define same night triggers for all objects with
at least a 25\% chance of being a supernova but no more than a 1\%
chance of being an asteroid. Discoveries worthy of same night follow-up
will be sent primarily to P60, although sources identified at the end
of the night may fall to Faulkes Telescope North (FTN; Haleakala, Hawaii) in rare cases where twilight preempts the
required observation. These requests will be issued as ToO triggers,
similar to the gamma ray burst triggers already handled by both P60 and FTN except
that interruption of integrations or read outs will not be required. 

After completing an observation, the data must be analyzed to extract
the target parameters (e.g. the magnitude and position in photometric
data or the spectral classification, redshift, etc. for spectral
observations). The responsibility for this data reduction lies with
the partner in charge of the instrument, and the high level data will
be sent back to the Marshal. The Marshal will then relay this
information back to the classifier to refine the classification, and
it will archive the observation in its own database for future
use.

\section{The First PTF Transient Detections}
\label{sec:sn}
We conclude with the first results from the PTF survey. PTF's first bona fide transient was detected at $\rm{m_{g^\prime}}$=18.62$\pm$0.04 mag on UT 2009 March 2.347 \citep{2009ATel.1964....1K} after subtraction of a reference template made from PTF images acquired on Feb 25, 26, and 28 (Figure \ref{fig:sn_img}). The transient, at RA = 14:23:55.82 Dec = +35:11:05.2 (J2000, uncertainty $<$1"), was offset 0.9" W and 2.7" S from the core of an SDSS galaxy with $\rm{m_g}$=17.78 and a redshift of z=0.0555.

We obtained spectra of the transient on March 8.388 UT with the Double Beam Spectrograph (DBSP) instrument on the  5.1-m Palomar Hale telescope. The transient turned out to be a normal Type Ia supernova at a redshift consistent with the apparent SDSS host. In this restframe, the expansion velocity derived from the minimum of the SiII (rest 635.5 nm) line was about 10,700 $\rm{km s^{-1}}$. The best match found with the Superfit SN spectral identification code \citep{Howell2005} was to SN 1994D at 8 days prior to maximum light. 

During the commissioning process many further transients have been detected and followed up (Table \ref{tab:march_transients}; Figure \ref{fig:sn_spect}). Eleven were detected during an R-band search of 498 square degrees \citep{ATel1983}, using reference images acquired by PTF prior to March 12. All eleven transients were spectroscopically classified on UT 2009 Mar 20 using DBSP. Twenty-five optical transients were found in a search of several nights of commissioning data taken between March 17 and March 28, photometrically followed up with P60, and spectroscopically classified using the Hale, Hobby-Eberly and Keck telescopes \citep{ATel2005}. A very bright supernova was detected on April 17 \citep{ATel2037} and a further 15 optical transients were discovered and followed up in May 2009 \citep{ATel2055}. These detections represent a small fraction of the ultimate PTF discovery power, as the majority of the commissioning time was used for acquiring reference images.  

These initial transient detections have opened the Palomar Transient Factory for business. The full PTF survey is planned to start in June 2009 and the project is scheduled to continue until at least the end of 2012.

\begin{table*}
\begin{center}
\caption{Optical Transients Discovered during PTF commissioning}
\label{tab:march_transients}
\begin{small}
\begin{tabular}{llllllllll}
\bf Name    &\bf  RA(J2000)    &\bf DEC(J2000)   &\bf Date UT 2009  &\bf  Mag   & \bf z&\multicolumn{2}{l}{\bf Offset from Host} & \bf Spect. Class.  \\ 
\hline 
PTF09a     & 14:23:55.82    & +35:11:05.2   & Mar 02.3470 & 18.6 & 0.06 & 0.9" W & 2.7" S & SN Ia \\
PTF09d & 09 15 36.57 & +50 34 51.5 & Mar 17.1476 & 19.5 &   &   \multicolumn{2}{c}{}        & CV   \\
PTF09e & 09 14 34.16 & +51 10 29.7 & Mar 17.1476 & 20.0 & 0.15  &   0.1"E& 0.8"S   & SN Ia\\  
PTF09f & 11 41 54.05 & +10 25 46.1 & Mar 17.1759 & 20.1 & 0.15  &   0.5"E &2.6"S   & SN Ia \\
PTF09g & 15 16 31.50 & +54 27 35.4 & Mar 17.2681 & 18.4 & 0.04  &   0.5"E &4.5"N   & SN II   \\
PTF09h & 08 00 47.28 & +46 56 53.8 & Mar 17.1418 & 20.2 & 0.12  &   0.4"E &0.4"S   & SN Ia \\
PTF09k  & 12 26 17.84 & +48 26 49.5 & Mar 17.1880 & 20.9 & 0.19 & 5.3"W & 6.3"N  & SN  Ia\\
PTF09o  & 04 05 02.73 & +73 24 54.2 & Mar 17.1343 & 18.5 &  &    \multicolumn{2}{c}{}         & CV?  \\
PTF09q  & 12 24 50.17 & +08 25 57.6 & Mar 17.2195 & 19.6 & 0.09  &   1.8"E& 3.9"S   & SN Ic  \\
PTF09r  & 14 18 58.68 & +35 23 16.1 & Mar 17.2572 & 19.2 & 0.03  &   0.7"E& 0.8"N   & SN II     \\
PTF09s  & 12 06 51.67 & +26 57 36.7 & Mar 17.1789 & 17.8 & 0.05  &   0.2"W& 0.0"N   & SN Ia   \\
PTF09t  & 14 15 43.29 & +16 11 59.1 & Mar 17.2664 & 18.6 & 0.04  &   4.6"E &1.4"S   & SN II      \\
PTF09u  & 14 29 39.61 & +39 09 32.8 & Mar 17.2557 & 20.0 & 0.13  &   0.0"W& 2.0"N   & SN Ia \\
PTF09v  & 13 27 10.50 & +31 30 32.5 & Mar 17.2091 & 19.2 & 0.12  &   0.7"W &0.5"N   & SN Ia  \\
PTF09x  & 13 21 45.15 & +42 33 06.2 & Mar 21.3851 & 20.2 & 0.25 & 0.8"W & 1.2"N  & SN Ia \\ 
PTF09y  & 11 16 34.27 & +03 32 02.8 & Mar 21.1860 & 20.1 & 0.19 & 0.2"W & 2.1"N & SN?\\  
PTF09z  & 11 54 42.23 & +55 18 10.7 & Mar 21.4275 & 20.1 & 0.19 & 0.1"W & 0.4"N & SN Ia  \\
PTF09aa & 11 33 20.71 & -09 24 40.3 & Mar 21.1961 & 19.0 & 0.12 & 0.0"W & 0.2"N & SN  Ia  \\
PTF09ab & 09 22 15.69 & +45 44 53.4 & Mar 21.3599 & 19.5 & 0.17 & 0.0"E & 0.1"N & SN Ia  \\
PTF09ac & 12 24 35.31 & +47 14 16.8 & Mar 21.4350 & 19.2 & 0.16 & 6.1"W & 1.2"N  & SN Ia  \\
PTF09ad & 11 03 06.64 & +50 09 36.3 & Mar 21.4313 & 19.8 & 0.20 & 0.6"W & 1.4"N & SN Ia  \\
PTF09aj & 09 45 30.46 & +06 32 25.0 & Mar 21.2155 & 17.8 & 0.09 & 0.2"W & 1.5"N  & SN Ia  \\
PTF09as & 12 59 15.85 & +27 16 41.3 & Mar 25.1642 & 19.1 & 0.19 & 1.7"W & 5.5"N & SN Ia  \\
PTF09av & 09 15 12.73 & +19 05 46.3 & Mar 25.1522 & 20.2 & 0.22 & 2.0"W & 2.9"S & SN Ia  \\
PTF09aw & 14 15 19.36 & +16 25 14.0 & Mar 25.2625 & 19.7 & 0.17 & 4.7"W & 4.5"S & SN Ia  \\
PTF09bc & 10 51 08.55 & +74 05 23.2 & Mar 20.2834 & 20.51 & 0.18 & 3.9"E & 1.9"N & SN Ia  \\
PTF09bd & 08 07 29.72 & +15 34 41.8 & Mar 20.1407 & 16.9 &  &     \multicolumn{2}{c}{}      & CV  \\
PTF09be & 14 10 18.54 & +16 53 38.8 & Mar 26.2304 & 19.0 &  & 4.7"E & 3.9"N   & SN II \\ 
PTF09bh & 12 24 39.20 & +08 55 59.2 & Mar 28.4155 & 20.5 & 0.18 & 1.6"W & 0.8"N &SN Ia\\  
PTF09bi & 11 46 50.12 & +11 47 55.3 & Mar 25.1613 & 19.6 & 0.11 & 2.6"W & 0.7"S &SN II?  \\
PTF09bj & 11 18 06.46 & +12 53 43.1 & Mar 25.1627 & 18.1 & 0.14 & 0.1"E & 1.3"N  & SN Ia  \\
PTF09bw & 15 05 01.97 & +48 40 03.9 & Mar 28.4039 & 20.3 & 0.15 & 0.8"W & 0.7"N & SN IIn?\\  
PTF09bx & 14 30 50.42 & +35 37 31.4 & Mar 26.2086 & 18.8 &   & 4.2"E & 3.3"S   & SN?  \\
PTF09by & 13 29 12.64 & +46 43 27.5 & Mar 26.2509 & 19.3 & 0.10  & 0.2"E & 0.9"N & SN Ia  \\
PTF09ct & 11 42 13.88 & +10 38 54.0 & Mar 27.1606 & 20.3 & 0.15 & 5.6"W & 5.2"N & SN II?  \\
PTF09cu & 13 15 23.15 & +46 25 09.4 & Mar 26.2509 & 17.9 & 0.06 & 7.5"W & 3.9"S & SN II  \\
PTF09dh & 14 44 42.08 & +49 43 44.9  & Apr 17.375 & 20.3 & 0.1 & &   &  SN      \\
PTF09ds & 14 09 16.65 & +53 06 13.2 & May 15.236 & 20.6 & 0.179 & 0.9"W&0.2"N  &  SN Ia      \\
PTF09ec & 13 12 54.49 & +43 28 36.0 & May 15.259 & 19.8 & 0.091 &  0.3"E&0.0"N  &  SN Ia \\
PTF09fb & 16 46 42.82 & +75 15 28.6 & May 16.198 & 17.7  & 0.042 &  0.0"E&0.2"S &  SN Ia    \\
PTF09fr & 14 50 00.12 & +44 55 05.8 & May 17.172 & 19.1  & 0.08 &  2.5"E&5.6"N  &  SN Ia     \\
PTF09fs & 17 36 44.28 & +53 40 12.3 & May 17.244 & 20.3 & 0.109 &  1.4"W&3.5"S  &  SN	     	\\     
PTF09fu & 16 33 10.88 & +53 05 30.4 & May 17.247 & 20.5 & 0.17 &  8.1"W&2.5"N   &  SN Ia     \\
PTF09go & 16 47 34.80 & +49 50 00.4 & May 17.237 & 18.1 & 0.047 &0.2"W&0.2"S  &  SN II	       \\
PTF09do & 17 38 26.04 & +53 23 00.4 & May 15.220 & 20.2 & 0.080 &  4.0"E&1.1"N  &  SN Ia    \\
PTF09ge & 14 57 03.10 & +49 36 40.8 & May 17.229 & 19.4 & 0.064 &  0.1"E&0.1"S  &  SN?	      	  \\   
PTF09gk & 15 06 11.06 & +53 17 42.9 & May 17.253 & 20.0  & 0.0 &   &  &  CV	     	    \\ 
PTF09gm & 15 27 48.59 & +41 35 34.0 & May 17.196 & 19.4 & 0.082 &  0.2"E&1.6"S  & SN Ia      \\
PTF09gn & 15 29 10.95 & +40 47 39.0 & May 17.196 & 20.4  & 0.139  &  1.1"W&0.4"S &  SN Ia      \\
PTF09ib & 15 44 38.77 & +45 47 51.0 & May 20.401 & 20.4 & 0.123  & 0.1"E&0.2"S &  SN Ia \\     
PTF09ij & 14 32 14.62 & +54 51 19.7 & May 20.288 & 20.3 & 0.123  & 3.7"W&2.1"N &  SN	\\       
PTF09jw & 12 54 23.74 & +56 43 57.5 & May 20.245 & 20.4 & 0.153  & 0.2"E&0.3"N &  SN Ia  \\ 
\end{tabular}
\end{small}
\end{center}
\end{table*}

\vspace{0.3cm}
\section*{Acknowledgments}
\small
This paper is based on observations obtained with the Samuel Oschin Telescope and the
60-inch Telescope at the Palomar Observatory as part of the Palomar
Transient Factory project, a scientific collaboration between the
California Institute of Technology, Columbia University, Las Cumbres
Observatory, the Lawrence Berkeley National Laboratory, the National
Energy Research Scientific Computing Center, the University of Oxford,
and the Weizmann Institute of Science. SRK and his group were partially supported by the NSF grant AST-0507734. J.S.B.  and  his group  were partially supported  by a Hellman Family
Grant, a  Sloan Foundation  Fellowship, NSF/DDDAS-TNRP grant CNS-0540352, and
a  continuing  grant  from  DOE/SciDAC.  TB, AP, WR and DAH are supported by the TABASGO foundation and the Las Cumbres Observatory Global Telescope Network  The  Weizmann
Institute PTF  partnership is supported  by an ISF equipment  grant to
AG.  AG's  activity is  further supported by  a Marie Curie  IRG grant
from  the EU,  and  by  the Minerva  Foundation,  Benoziyo Center  for
Astrophysics, a research grant  from Peter and Patricia Gruber Awards,
and the  William Z.  and  Eda Bess Novick  New Scientists Fund  at the
Weizmann  Institute.  EOO  thanks partial  supports from  NASA through
grants    HST-GO-11104.01-A;     NNX08AM04G;    07-GLAST1-0023;    and
HST-AR-11766.01-A.   A.V.F.  and  his group  are grateful  for funding
from NSF  grant AST--0607485, DOE/SciDAC  grant DE-FC02-06ER41453, DOE
grant  DE-FG02-08ER41563,  the TABASGO  Foundation,  Gary and  Cynthia
Bengier, and the Sylvia and Jim Katzman Foundation. S.G.D. and A.A.M. were supported in part by NSF  grants  AST-0407448  and   CNS-0540369, and also by the Ajax Foundation. The National Energy Research Scientific Computing Center, which is supported by the Office of Science of the U.S. Department of Energy under Contract No. DE-AC02-05CH11231, has provided resources
for this project by supporting staff and providing computational resources and data storage. 
 A.V.F. and his group are grateful for funding from NSF grant AST--0607485, DOE/SciDAC grant DE-FC02-06ER41453, DOE
grant  DE-FG02-08ER41563, the TABASGO Foundation, Gary and Cynthia
Bengier, the Richard and Rhoda Goldman Fund, and the Sylvia and
Jim Katzman Foundation.  L.B.'s research is supported by the NSF via grants PHY 05-51164 and AST 07-07633. MS acknowledges support from the Royal Society and
the University of Oxford Fell Fund.

\pagebreak

\bibliographystyle{apj}

\bibliography{refs}

\label{lastpage}

\end{document}